\documentclass[]{aa}  

\usepackage{epsfig}
\usepackage{natbib} %Biblography
\bibpunct{(}{)}{;}{a}{}{,} % to follow the A&A style
\usepackage{graphicx}
\begin{document}
   \title{Photospheric downward plasma motions in the quiet-Sun}

   %\subtitle{I. Overviewing the $\kappa$-mechanism}

   \author{C. Quintero Noda \inst{1,2}
          \and
          B. Ruiz Cobo\inst{1,2}	  
          \and
	      D. Orozco Su\'arez\inst{1,2}
          }

   \institute{Instituto de Astrof\'isica de Canarias, E-38200, La Laguna, Tenerife, Spain.\ \email{cqn@iac.es}
	\and
         Departamento de Astrof\'isica, Univ. de La Laguna, La Laguna, Tenerife, E-38205, Spain
}

   \date{Received January 2014 / Accepted March 2014} 
% \abstract{}{}{}{}{} 
% 5 {} token are mandatory
  \abstract
   {We analyze spectropolarimetric data taken with the \textit{Hinode} spacecraft in quiet solar regions at the disk center. Distorted redshifted Stokes $V$ profiles are found showing a characteristic evolution that always follows the same sequence of phases.}
   {We aim to characterize the statistical properties of these events and recover the stratification of the relevant physical quantities to understand the nature of the mechanism behind them.}
   {We have studied the statistical properties of these events using spectropolarimetric data from \textit{Hinode}/SP. We also examined the upper photosphere and the low chromosphere using Mg~{\sc i} b$_2$ and Ca~{\sc ii h} data from \textit{Hinode}. Finally, we have applied the SIRGAUSS inversion code to the polarimetric data in order to infer the atmospheric stratification of the physical parameters. We have also obtained these physical parameters taking into account dynamical terms in the equation of motion.}
   {The Stokes $V$ profiles display a bump that evolves in four different time steps, and the total process lasts 108 seconds. The Stokes $I$ shows a strongly bent red wing and the continuum signal exhibits a bright point inside an intergranular lane. This bright point is correlated with a strong redshift in the Mg~{\sc i} b$_2$ line and a bright feature in Ca~{\sc ii h} images. The model obtained from the inversion of the Stokes profiles is hotter than the average quiet-Sun model, with a vertical magnetic field configuration and field strengths in the range of kG values. It also presents a LOS velocity stratification with a Gaussian perturbation whose center is moving to deeper layers with time. The Gaussian perturbation is also found in the gas pressure and density stratification obtained taking into account dynamical terms in the equation of motion.}
   {We have examined a particular type of event that can be described as a \textit{plasmoid} of hot plasma that is moving downward from the top of the photosphere, placed over intergranular lanes and always related to strong magnetic field concentrations. We argue that the origin of this plasmoid could be a magnetic reconnection that is taking place in the chromosphere.}
   %{}

\titlerunning{Photospheric downward plasma motions in the quiet-Sun}          
                             
   % \titlerunning                          
                             
\authorrunning{Quintero Noda}
%\titilerunning

   \maketitle
   
%\authorrunning

%________________________________________________________________
\section{Introduction}

Understanding of quiet-Sun magnetism has improved enormously in recent years, mainly thanks to the successful performance of several instruments operating beyond the Earth's atmosphere, such as the spectropolarimeter \citep{Lites2013_SP} on board the \textit{Hinode} \citep{Kosugi2007,Tsuneta2008} and IMaX \citep{IMaX} on board \textit{SUNRISE} \citep{Barthol2011,Solanki2010}. These instruments have provided data with spatial resolutions comparable to, or better than, those available in the past, see, for example, SST \citep{Scharmer2003}, and very significantly improved temporal consistency, going beyond the typical granular evolutionary time scales. One of the most interesting features in the quiet-Sun photosphere is the high speed magnetized flows that can be deduced from highly distorted Stokes $V$ profiles detected in different solar regions \citep{Shimizu2008}. These profiles present a strongly redshifted signal and a multiple lobe structure. However, in spite of the efforts of previous authors, the origin of these strong flows and their length scale in the solar atmosphere still remain a subject for debate.

One of the proposed responsible mechanisms is a convective instability inside a magnetic flux tube. The convective collapse process was first suggested on theoretical grounds by \citet{Parker1978}, \citet{Webb1978} and \citet{Spruit1979}. Recent studies  have tentatively identified this mechanism through observations \citep{BellotRubio2001,Nagata2008,Fischer2009} as the amplification of the magnetic field induced by a strong downflow, with line-of-sight (LOS) component speeds of 6 km s$^{-1}$.  It has also been proposed that these downflows hit dense layers and rebound as supersonic upflows \citep{Grossmann1998}. These upflows would explain the extremely blueshifted signals observed by \citet{SocasNavarro2005}. Another possible mechanism that could produce the strong flows is  magnetic reconnection \citep{Parker1963}. This mechanism was proposed by \cite{Borrero2013} to explain the photospheric stratification obtained through the inversion of the strongly Doppler-shifted profiles first reported in \cite{Borrero2010}.

Chromospheric layers also present ubiquitous high speed plasma events in form of needle-like structures called spicules that appear in limb observations \citep{Beckers1968}. The recent rediscovery of a new type of spicule  \citep[type {\sc ii} spicules,][]{DePontieu2007} with shorter lifetimes (10--150 s), smaller diameters ($<$200 km compared to $<$500 km for type {\sc i} spicules), and faster rise times confirms the complex nature of these solar structures, which are still far from being well understood. The Solar disk center counterpart of the limb spicules was first studied by \cite{Langangen2008} and then by \cite{Sekse2012,Sekse2013}, who revealed the existence of highly blueshifted signals in the chromospheric Ca {\sc ii} line, named by the first author rapid blueshifted excursions (RBEs). They are characterized by a length of about 1.2 Mm and a width of 0.5 Mm. The RBEs appear to be located near strong magnetic concentrations (network patches), but not on top of them. The Doppler shifts associated with the RBEs suggest velocities of 15--20 km/s. Owing to the highly energetic nature of RBEs, previous authors conclude that the asymmetric blueward spectral profile has to be generated as the result of a magnetic reconnection process that occurs in the chromosphere.

In this paper we examine the polarimetric properties of Fe {\sc i} Stokes $V$ profiles from \textit{Hinode}/SP that display a strong Doppler-shifted signal on the red side of the line. We then follow their vertical trace through an  analysis of the upper photospheric Mg~{\sc i} b$_2$ line and chromospheric Ca {\sc ii h} spectral band. Finally, we infer the atmospheric stratification of the physical parameters using the SIRGAUSS inversion code. The aim of the study is to reveal the nature of these strong Doppler-shifted signals and the physical mechanism that produces them.

\section{Observations and data analysis}

\begin{figure*}
\centering
\includegraphics[width=18cm]{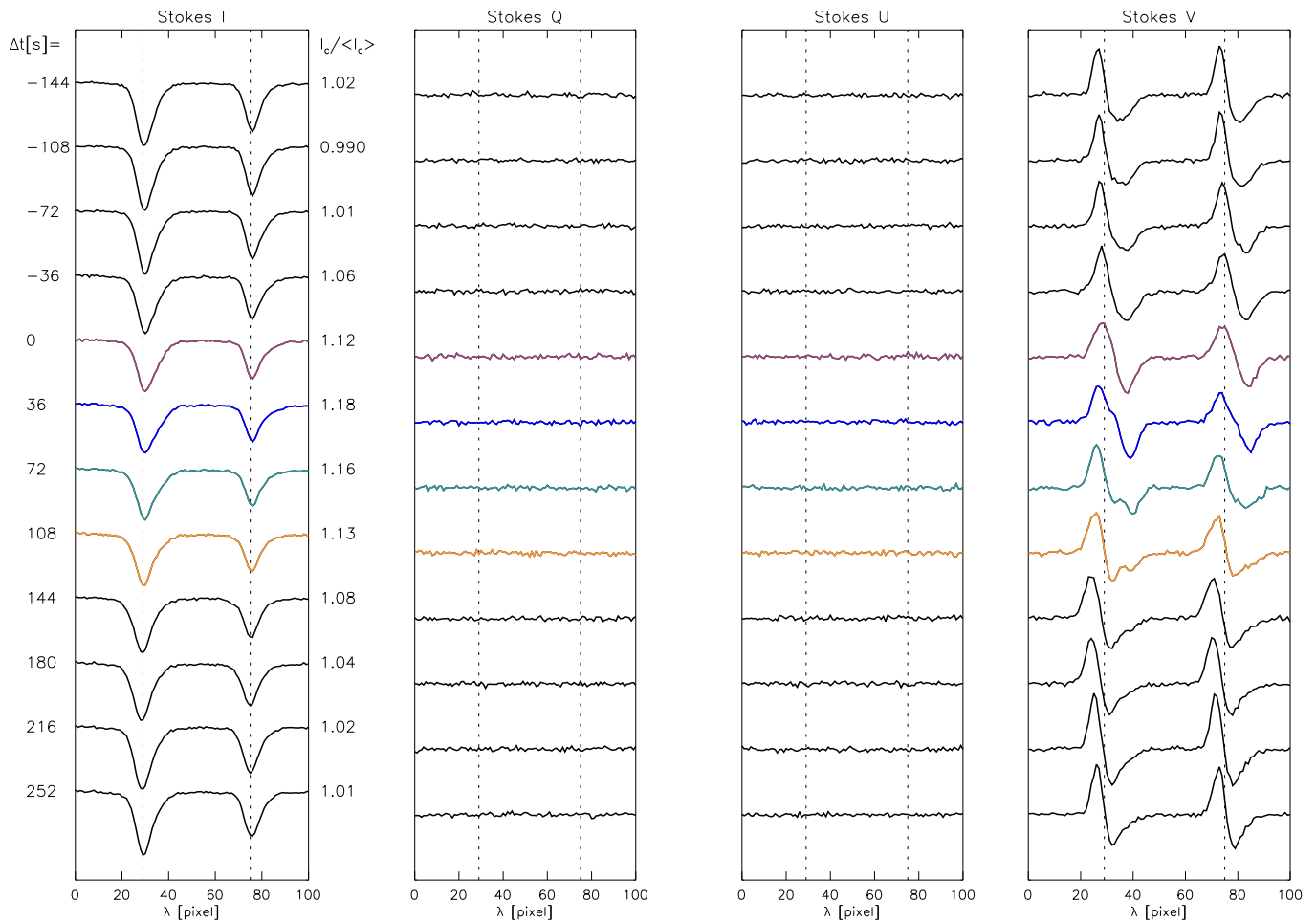}
\caption{Temporal evolution of a single pixel from one of the 11 detected cases in the SP time series. Each column corresponds to a different Stokes parameter and time runs from top to bottom. Stokes $Q$ and $U$ are always below the noise. Stokes $V$ shows a bump at the top-center of the line (purple; $\Delta t=0$) that apparently moves, crossing the center of the line to the red wing (orange; $\Delta t=108$) and finally vanishes. The appearance of the bump coincides with an enhancement of the continuum intensity signal (values on right of the first column represent the continuum signal divided by the mean continuum intensity of the whole map). After the disappearance of the bump the continuum intensity decreases.}
\label{figper1}
\end{figure*}

The polarimetric data used in this paper were obtained with the spectropolarimeter (SP) on board \textit{Hinode}. In particular, we took two different data sets: one consisting of two scans, also called \textit{normal} maps, of $328^{\prime\prime}\times154^{\prime\prime}$ recorded on 2007 March 10$^{\rm th}$ and 2007  October 15$^{\rm th}$ pointing at the disk center. The SP instrument measures the Stokes spectra of the Fe~{\sc i} 6301.5 and 6302.5 \AA \ lines with a spectral sampling of 2.15 pm pixel$^{-1}$ and a spatial sampling of 0.16$^{\prime\prime}$. The exposure time is 4.8 seconds per slit position making it possible to achieve a noise level of $1.1\times10^{-3}I_{c}$ in Stokes $V$ and $1.2\times10^{-3}I_{c}$ in Stokes $Q$ and $U$, where I$_{c}$ stands for the continuum intensity.

The second data set is a time series observation, based on a raster scan of 18 scanning steps with 1.6 seconds of exposure time per slit position, spanning 3 hours of observation on 2007  September 25$^{\rm th}$. In this case, the field of view is $2.9^{\prime\prime}\times38^{\prime\prime}$ and the final time cadence is 36 seconds. These observations belong to the \textit{Hinode} Operation Plan 14, entitled \textit{Hinode/Canary Islands campaign}, and allow us to analyze the evolution of small magnetic structures present in the quiet-Sun although the noise level is worse in comparison with the first data set.

The \textit{Hinode} Broadband Filter Imager (BFI) \citep{Tsuneta2008} instrument acquired simultaneous images of the photosphere in the CN band head at 388.3 nm (filter width of 0.52 nm) and of the chromosphere in the Ca~{\sc ii h} line at 396.85 nm (filter width of 0.22 nm). The exposure times were 0.1 s and 0.3 s, respectively. The BFI covered a region of $19.12^{\prime\prime}\times74.1^{\prime\prime}$ with a pixel size of $0.055^{\prime\prime}$ and took images with a cadence of 30 seconds.

We also have data from the Narrowband Filter Imager (NFI) consisting of shutterless Stokes $I$ and $V$ images in the red and blue wings of the Mg~{\sc i} b$_2$ line at 517.27 nm. From these data we constructed magnetograms and Dopplergrams as

\begin{equation}
M=\frac{1}{2}\left(\frac{V_b}{I_b}-\frac{V_r}{I_r}\right),
\end{equation}

\begin{equation}
D=\frac{I_b-I_r}{I_b+I_r},
\end{equation}
where \textit{r} and \textit{b} indexes stand for the red and blue sides of the line.

Data from CN and Ca~{\sc ii h} spectral bands are a series of images of the same FOV, taken every 30 seconds. Due to the difference between the cadence of the CN and Ca~{\sc ii h} images and the cadence of each \textit{Hinode}/SP map (36 seconds), we have chosen the closest image of the CN and Ca~{\sc ii h} bands to the SP map snapshot. We selected the time reference from the central slit of the SP map (slit 9) and chose the image of CN or Ca~{\sc ii h} with the observing time closest to the ninth slit time. The broad-band images that we have chosen come before the slit acquisition time or after it, but the temporal difference between them was never larger than 18 seconds, which corresponds to half of the time cadence of each \textit{Hinode}/SP map.

For the Mg~{\sc i} b$_2$ data we have followed the same method as for the Ca~{\sc ii h} images but in this case the field of view of the Mg~{\sc i} b$_2$ is larger than the FOV of the \textit{Hinode}/SP maps. We have selected the region of the magnesium image that corresponds to the \textit{Hinode}/SP raster scan map to analyze the possible relationship between them using strong circular polarization signals on the Mg~{\sc i} b$_2$ magnetograms as reference.

The events of interest in this study are those representing a strong Doppler-shifted Stokes $V$ signal. We have followed the method presented by \cite{MartinezPillet2011a} to find them, examining the Stokes $V$ signal at $\pm 272$~m\AA \ from the center of the Fe {\sc i} 6302.5 \AA \ line and generating \textit{far wing} magnetograms from the red and the blue sides of that line. We then analyzed these magnetograms looking for pixels with Stokes $V$ signal above 0.5\% of the mean continuum intensity of the map (hereafter I$_c$). As the detected events were grouped in patches of different sizes, we decided to define a minimum area of detection of four pixels to avoid the contribution of isolated bad pixels.

A large number of redshifted patches compared to blueshifted ones was found. We have detected a family of events that show a characteristic evolution of their Stokes $V$ profile shapes inside these redshifted cases. We analyzed their evolution using the raster scan observing mode, studying at the same time the information available from higher layers where the Mg~{\sc i} b$_2$ and Ca~{\sc ii h} lines are formed.

Finally, the atmospheric models of these events were inferred using the SIRGAUSS inversion code \citep{BellotRubio2003}. This code is based on SIR \citep[Stokes inversion based on Response Functions;][]{RuizCobo1992} and considers Gaussian perturbations along the LOS stratification for the atmospheric parameters.

\begin{figure*}
\centering
\hspace*{-0.3cm}\includegraphics[width=20cm]{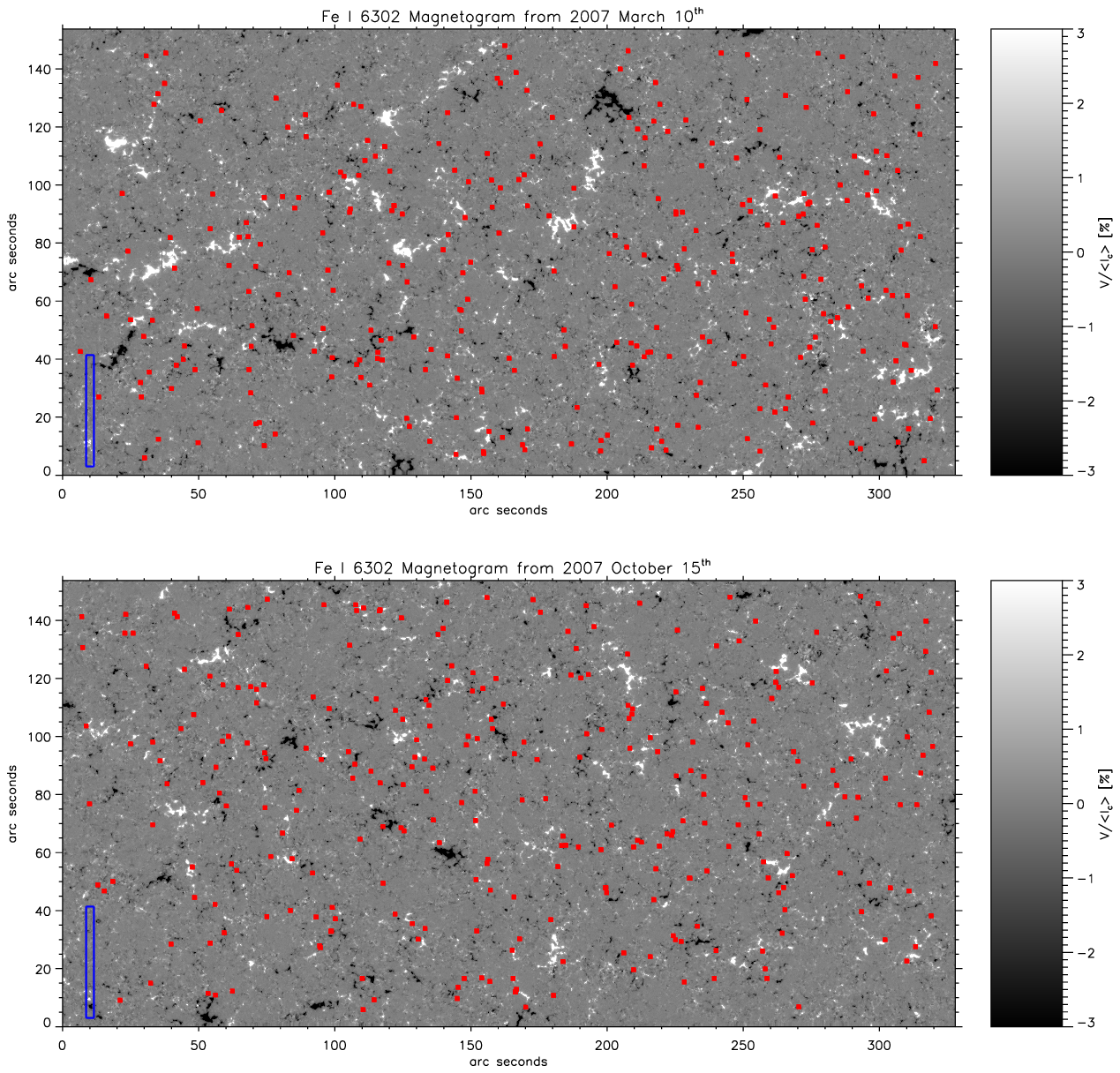}
\caption{Location of the redshifted events (red filled squares) over the Fe~{\sc i} 6302.5 \AA \ magnetogram from the \textit{normal} maps recorded on 2007 March 10$^{\rm th}$ (upper panel) and 2007 October 15$^{\rm th}$ (bottom panel). The size of the filled squares is greater than the size of the events in order to make them visible. The left blue rectangle represents the FOV of the raster scan mode.}
\label{figmagn}
\end{figure*}

\section{General properties of the redshifted events}\label{general_properties}

\subsection{Polarimetric characteristics and evolution}\label{amplitudes}

The detected redshifted events appear always at intergranular lanes. The Stokes profiles change their shape as Fig.~\ref{figper1} shows. Each column corresponds to the evolution of the Stokes profiles from a single pixel in one event. The temporal cadence is 36 seconds. We have marked with colors the most important part of the event that starts at $\Delta t=0$. Before the beginning of the process, Stokes $V$ is strongly redshifted and, in addition, presents a continuum enhancement in Stokes $I$ that starts at $\Delta t=-72$. The Stokes $V$ profile then suddenly changes its shape (purple; $\Delta t=0$); a small bump appears on the top part of the profile at the zero crossing point (blue; $\Delta t=36$); then, it shifts to the red wing, distorting the Stokes $V$ red lobe (green; $\Delta t=72$); and finally it appears at the red wing of the Stokes $V$ profile (orange; $\Delta t=108$). After that instant, the Stokes $I$ continuum signal relaxes to values closer to those displayed by quiet-Sun intergralunar lanes and the Stokes $V$ profiles present an extended red wing and evolve in two different ways: the profiles could recover a typical antisymmetric Stokes $V$ shape with the absence of redshifted signal or, the most common behaviour, the polarization signals vanish.

The Stokes $V$ profiles show a strong redshifted signal during the whole process. Although the cores of Stokes $I$ (first column of Fig.~\ref{figper1}) are not highly redshifted, the red wings are strongly distorted. The Doppler shift and the continuum enhance mentioned before start to decrease during the late steps (green and orange). Remarkably, the Stokes $V$ amplitude of the Fe~{\sc i} 6301.5 \AA \ line is always higher than the amplitude of 6302.5 \AA , its neighbouring Fe {\sc i} line.  This is an unexpected attribute of the Stokes $V$ profiles of these events because in the weak field regime the first line, Fe~{\sc i} 6301.5 \AA \ (the less sensitive one to the magnetic field line), always has  a lower Stokes $V$ amplitude. In addition, although the wavelength location of the bump inside each Stokes $V$ profiles is the same for both lines, the bump is more prominent in the first line, i.e.\ Fe~{\sc i} 6301.5 \AA.

In the region where the event occurs, Stokes $V$ amplitudes are usually between 5\% and 10\% of the mean continuum signal ($I_{c}$), whereas $Q$ and $U$ are always below the noise level (see Fig.~\ref{figper1}), indicating that the transverse component of the magnetic field has to be small, i.e.\ the magnetic structure is nearly vertical. Moreover, neither structures presenting opposite polarities nor linear polarization signals appear to be close to the region where the event evolves.

Regarding the Stokes $I$, it is important to note that the event appears in an intergranular lane, where continuum intensities in the quiet-Sun have values in the range 0.8--0.95 $I_c$, while the pixels that display strong Stokes $V$ redshifted signals have Stokes $I$ continuum intensity values over 1.1 $I_c$. For this reason, this process always comes into view as a bright point in the continuum map. In addition, if we define the line depth of Stokes $I$ as the difference between the intensity of the continuum and the core of the line, we find that the Stokes $I$ profiles have a small line depth in both lines, reaching values in the range 0.4--0.5 $I_c$ while typical quiet-Sun profiles usually present values of 0.7--0.8 $I_c$. This low line depth value is due to a significant increase of the line core intensity that is greater than the enhancement of the continuum signal.

\begin{figure*}
\centering
\hspace*{-0.8cm}\includegraphics[width=18.6cm]{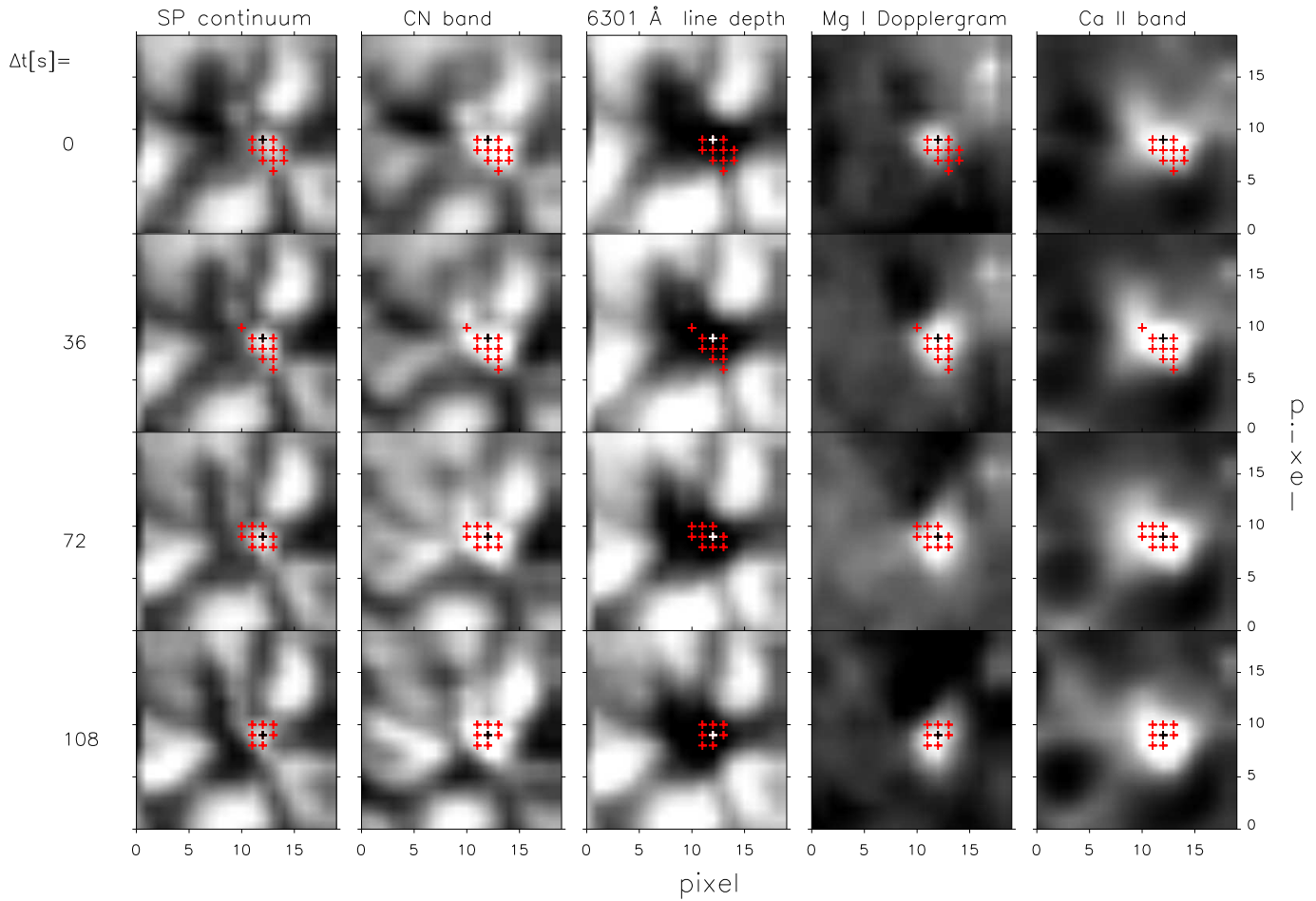}
\caption{The first column corresponds to the \textit{Hinode}/SP continuum signal. The second one is the CN spectral band. The third column shows the line depth of the Fe~{\sc i} 6301.5~\AA \ line. The fourth column presents the Mg~{\sc i} b$_2$ Dopplergram signal (with downflows in white and upflows in black),  and the last one shows the Ca~{\sc ii h} intensity maps. Red crosses correspond to the locations of the pixels that are highly redshifted. The time interval runs from top to bottom and is the depicted by colors in Fig.~\ref{figper1}. The pixel analyzed in that figure is marked with a black cross (white in the third column). We have reduced the spatial resolution of CN, Ca~{\sc ii h} and Mg~{\sc i} b$_2$ images to match the \textit{Hinode}/SP spatial resolution.}
\label{figcal}
\end{figure*}

\subsection{Lifetime, size and rate of occurrence}

We detected 11 cases that show the same behaviour as the one presented in Fig.~\ref{figper1} during the three hours that the time series lasts. A closer examination of their evolution allows the determination of a mean life time of 360$\pm$74 seconds. To calculate the lifetime of each event we have assumed that the initial stage corresponds to the simultaneous detection of the strongly redshifted signal in Stokes $V$ and the enhancement of the continuum Stokes $I$ signal while the last stage corresponds with the disappearance of the strongly redshifted Stokes $V$ signal and the relaxation of the Stokes $I$ continuum signal. The different positions of the bump displayed by the Stokes $V$ profile are indicated by colors in Figure \ref{figper1}. We always detected a different bump step in consecutive images, which indicates that the lifetime of each step has to be less than the cadence of the observing mode, i.e.\ 36 seconds. When we examined the Stokes $V$ signals from adjacent pixels inside the event for a single snapshot, we found the coexistence of one or two different and consecutive bump stages, i.e.\ different colors in Figure \ref{figper1}. This means that more than one bump step could be found inside the magnetized patch whose mean surface occupies 4$\pm$1.4 pixels. We also noticed that the events do not stay in the same place but move horizontally inside an 0.5 arcsec$^2$ area.

In order to establish a rate of occurrence, we examined the large FOV of the \textit{normal} maps. We looked for profiles that present the same polarimetric characteristics explained in the previous section. To analyze this enormous amount of data, we plotted a $5\times5$ pixel panel centred  on the cases that show a strongly redshifted signal at 272 m\AA \ from the Stokes $V$ center. After visual inspection of each panel, we selected the events that have a spatial distribution similar to the cases examined using the time series observation. The total number of detected events was 302 for the map recorded on 2007 March 10$^{\rm th}$ and 290 for the map of 2007 October 15$^{\rm th}$. The continuum signal reveals that these events, as in the time series, appear as bright features in intergranular lanes. Concerning the distribution of these profiles with respect to the magnetic field topology, Fig.~\ref{figmagn} shows the locations of the center of each event by red filled squares over the magnetogram maps obtained from the Fe~{\sc i} 6302.5~\AA \ line for both dates. The distribution of the events reveals that they are located in the surroundings of supergranular cells related to the strong magnetic activity represented by the network regions. However, the events never appear on top of these network regions.

The polarimetric analysis done in the previous section revealed that the different bump phases, represented by colors in Fig.~\ref{figper1}, last more than 100 seconds. This lifetime is much longer than the exposure time of each slit position used to obtain the maps of Fig.~\ref{figmagn}, i.e.\ $4.8$ s, allowing us to consider the magnetograms as single snapshots. In order to find the rate of occurrence of these events we need to divide the mean number of cases detected in both maps (296 $\pm$ 8.4) by the area of one map ($328^{\prime\prime}\times154^{\prime\prime}$). The resulting rate of occurrence of $5.9\times10^{-3}$ cases per arcsec$^{2}$ indicates that we would find nearly 60 different events in a field of view of 100$\times$100 arcsec$^{2}$. If we compare these properties with a typical granule structure (size of 1.5\arcsec and lifetime of 10 min) we can establish that these events are an uncommon, small and transient phenomenon that occur inside of intergranular lane regions.

\subsection{Chromospheric response}

Figure \ref{figcal} shows the Ca {\sc ii h}, CN and Mg~{\sc i} b$_2$ data. Time runs from top to bottom. The first column is the SP continuum signal. The second column is the CN spectral band. The third column is the line depth of the Fe~{\sc i} 6301.5 \AA \ line. The fourth column displays the Mg~{\sc i} b$_2$ Dopplergram and the fifth column shows the Ca~{\sc ii h} images. We have interpolated these maps to the SP spatial resolution to compare the CN, Ca~{\sc ii h} and Mg~{\sc i} b$_2$ images with the \textit{Hinode}/SP observations. Each row corresponds to a different time, running from top to bottom. We have chosen the same time steps we plotted in different colors in Fig.~\ref{figper1} starting at $\Delta t=0$. Red crosses indicate the position of the pixels of the selected event, i.e.\ pixels with enough signal at 272 m\AA. Additionally, we have marked with a black (or white in the third column) cross in each image the individual pixel displayed in Figure \ref{figper1}.

The continuum map shows that the event always appears over a bright region inside an intergranular lane. The CN band presents the same properties, displaying a bright feature at the positions marked by the red crosses. The next column shows the line depth; a black patch means that the line depth is below $0.5$ $I_c$, whereas a white patch indicates that the line depth is larger than $0.8$ $I_c$. The event always occurs inside a black region, meaning that the line depth is very small during the process. As we mention in Section \ref{amplitudes}, the enhancement of the line core of the 6301.5~\AA \ is large enough to produce a low Stokes $I$ line depth even when the continuum signal is above 1.1 I$_c$. This strong Stokes $I$ core enhancement complements the information of the previous section, where we noted the larger Stokes $V$ amplitude of this line, and indicates that the upper photosphere is where the process forms.

Magnesium Dopplergrams show a strong downflow patch in the same region where the event is detected. This patch with large downflows starts before, or at the same time as, the detection of the events by SP lines for the 11 different cases detected. We have omitted the magnesium magnetogram because it shows the same information as the Fe~{\sc i} 6302.5~\AA \ magnetogram: a unipolar patch isolated from opposite polarity regions. Finally, the Ca~{\sc ii h} images show a bright feature where the pixels of the event are plotted, suggesting that these Fe~{\sc i} distorted profiles have a counterpart in the region of formation of the Ca~{\sc ii h} line.

The correlation between photospheric downflows and bright features in the Ca~{\sc ii h} line has been mentioned before by \cite{Shimizu2008} and \cite{Fischer2009}. The latter work revealed an extensive variation in the Ca~{\sc ii h} signal during the evolution of the process. We also found this large variation in the region where the event occurs. Due to this substantial intensity fluctuation and the small field of view of the time series we have no clear hints to establish whether the bright Ca~{\sc ii h} feature was there before the detection of the event by the iron lines or whether it started at the same time. We can establish only that all the cases show a correlated bright point in Ca~{\sc ii h} data, and that this bright point never appears after the detection of the bump in the iron lines.

\begin{figure*}
\centering
\hspace*{-1.4cm}\includegraphics[width=20.0cm]{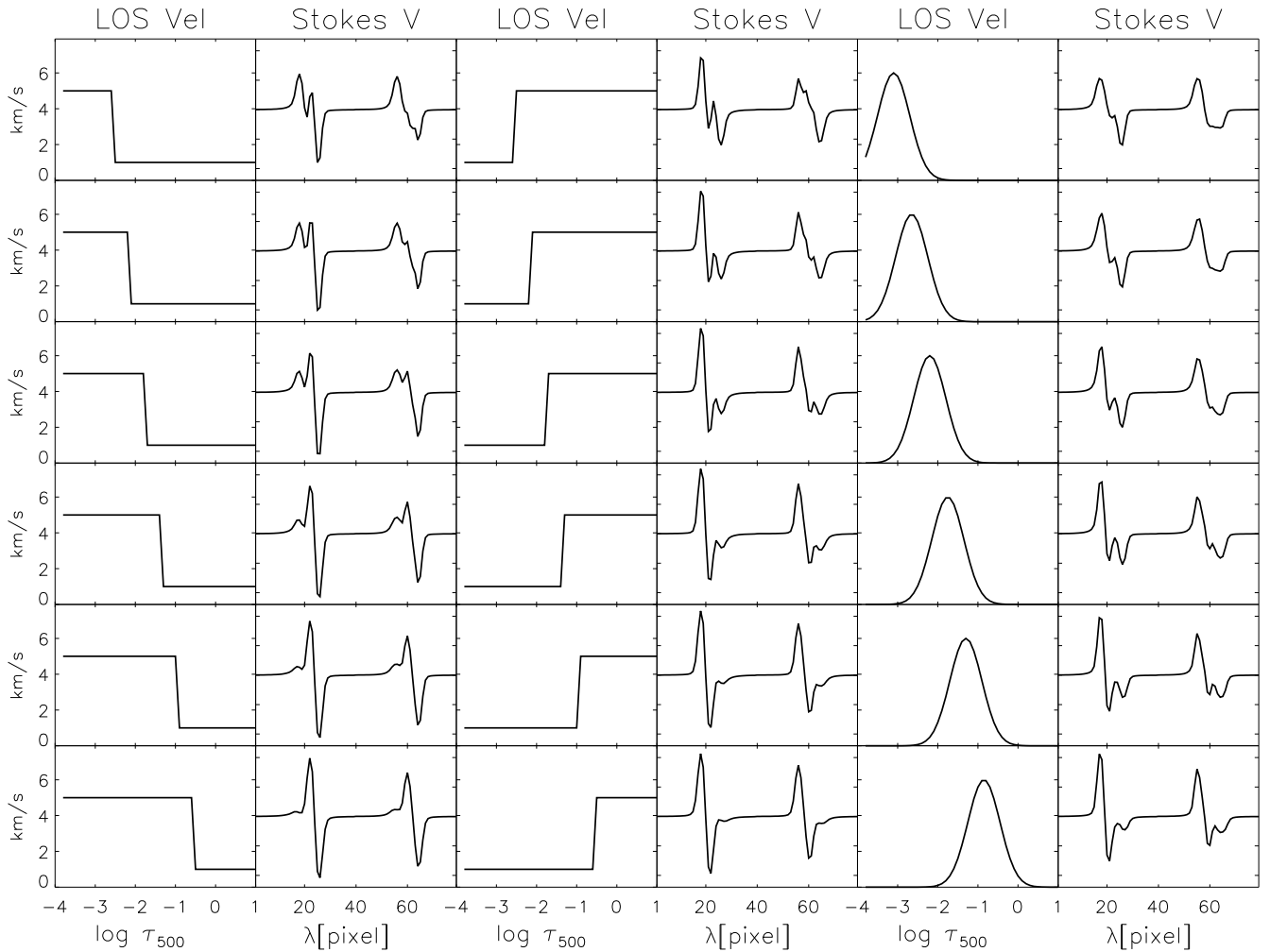}
\caption{Each pair of columns shows the LOS velocity stratification and the corresponding synthesized Stokes $V$ profiles. The temperature of the model used in the synthesis is hotter than the HSRA reference atmosphere. The magnetic field vector has a constant 1 kG intensity and zero degrees in inclination and azimuth. The first column shows a LOS velocity jump of 5 km/s that moves from the top to the bottom of the photospheric layers. The second column shows the resulting Stokes $V$ profiles. The third column shows an opposite jump, having higher velocities at lower heights. The Stokes $V$ profiles obtained from this velocity configuration are plotted in the fourth column. The last column shows the synthetic profiles when the LOS velocity stratification has a Gaussian perturbation, as shown in the fifth column.}
\label{figsint}
\end{figure*}

\section{Forward modelling}

We can obtain physical information on the mechanism that produces these events by the inversion of the Stokes profiles using the SIR code. Due to the high freedom in the modelling of the atmosphere allowed by the code it is advisable, before starting with the inversion of the Stokes profiles, to build a model whose profiles qualitatively show the same distortion as the observed ones.

After a large number of tests, we concluded that the bumps displayed by the observed Stokes $V$ profiles can be well reproduced using a model atmosphere that contains a strong jump in its LOS velocity stratification. However, the perturbation bump is not static in one position on the Stokes $V$ profiles when time evolves: it appears in the middle of the line, in the red lobe and finally in the red wing. For this reason the synthetic profiles need different jump stratifications to reproduce the different locations of this bump.

Figure \ref{figsint} shows different examples of the Stokes $V$ profiles that can be obtained using an ideal jump or a Gaussian profile in the LOS velocity stratification. The figure is divided into columns, with the LOS velocity stratification first and then the synthesized Stokes $V$ profile.

The amplitude of the Stokes $V$ profile for the Fe~{\sc i} 6301.5 \AA \ line is always lower than that of the second line when we use the temperature stratification from a quiet-Sun model such as the HSRA \citep{Gingerich1971}. The easiest way we found to change the Stokes $V$ amplitude behavior was to choose an atmospheric model hotter than that given by the HSRA reference atmosphere model. The difference between temperature stratifications is larger in the region of line formation. With this atmospheric configuration we could reproduce the observed amplitude of Fe~{\sc i} 6301.5 \AA \ during the whole process. This hotter atmospheric model also reproduces the increase in the signal level of the line core observed in the previous section. For the magnetic field configuration we chose a vertical and constant magnetic field. We selected a constant field intensity through the atmosphere owing to the absence of significant area asymmetry in the Stokes $V$ profiles, and a vertical magnetic inclination because the Stokes $Q$ and $U$ signals are always below the noise.

The first column of Figure \ref{figsint}  shows the first LOS velocity configuration, a jump of 5 km/s from top to bottom of the atmosphere, while the second column shows the synthetic and peculiar Stokes $V$ profiles that this velocity stratification produces. We discarded this solution because it produces a bump in the blue lobe, something that never happens in the observational data.

The fourth column shows the Stokes $V$ profiles coming from a LOS velocity configuration with a jump of 5 km/s from the bottom to the top of the atmosphere, i.e. a LOS velocity component that is only present at the base of the photosphere. In this case, the Stokes $V$ profiles resemble the observed ones, and the evolution of the bump is reproduced as the jump goes through different heights. However, in the polarimetric study in Section \ref{amplitudes} we pointed out that the wavelength location of the bump seems to be the same for both lines, something that is not reproduced in the first three rows of this column. We have never detected this apparent mismatch between the location of the bump in both lines in the observations (see Figure \ref{figper1}).

Finally, the fifth column shows a Gaussian perturbation whose center moves from the top to the bottom of the atmosphere. This LOS velocity stratification reproduces the displacement of the bump from the center of the line to the red wing, as the real profiles show. In this case the position of the bump seems to be almost the same for both lines, although it is sometimes not well defined in the second line, Fe~{\sc i} 6302.5 \AA. It should be noted that a Gaussian perturbation with a very large width will produce the same effect on the profiles as the LOS velocity model displayed in the third column owing to the small optical depth range of sensitivity for the two iron lines examined. This property implies that for the synthesis of the Stokes profiles a jump or a Gaussian with a large width will give the same results.

Due to the similarities between the real and synthetic profiles displayed in the sixth column of Fig.~\ref{figsint}, we decided to employ the model with the Gaussian perturbation in the LOS velocity to invert the profiles. We have used the SIRGAUSS inversion code \citep{BellotRubio2003} instead of SIR to obtain the atmosphere stratifications from the inversions of the Stokes profiles.

\section{Stokes profiles inversions}\label{Sirprocedure}

The SIRGAUSS code is a variation of the original SIR program that has as free parameters those corresponding to the background atmosphere together with the parameters needed to define a Gaussian perturbation for each physical quantity. The Gaussian perturbation profile is the same for all of the physical quantities, i.e.\ the same width and center, allowing a different amplitude for each physical quantity.

The configuration for the inversion code is based in a two-component model with a non-magnetic atmosphere and a magnetic atmosphere. The non-magnetic atmosphere has freedom only in temperature and LOS velocity. The magnetic component uses a Gaussian perturbation in LOS velocity and temperature  stratifications. The background under the Gaussian perturbation is kept constant to 0 km/s for the LOS velocity while it is inverted for the temperature stratification. The magnetic field, the microturbulence and the macroturbulence are constant with height, but their value can change. Inclination and azimuth angles are fixed to 0 degrees. Finally, the filling factor between components is also inverted during the process. As a result, the total number of free parameters is 12.

\begin{figure*}
\centering
\includegraphics[width=18.5cm]{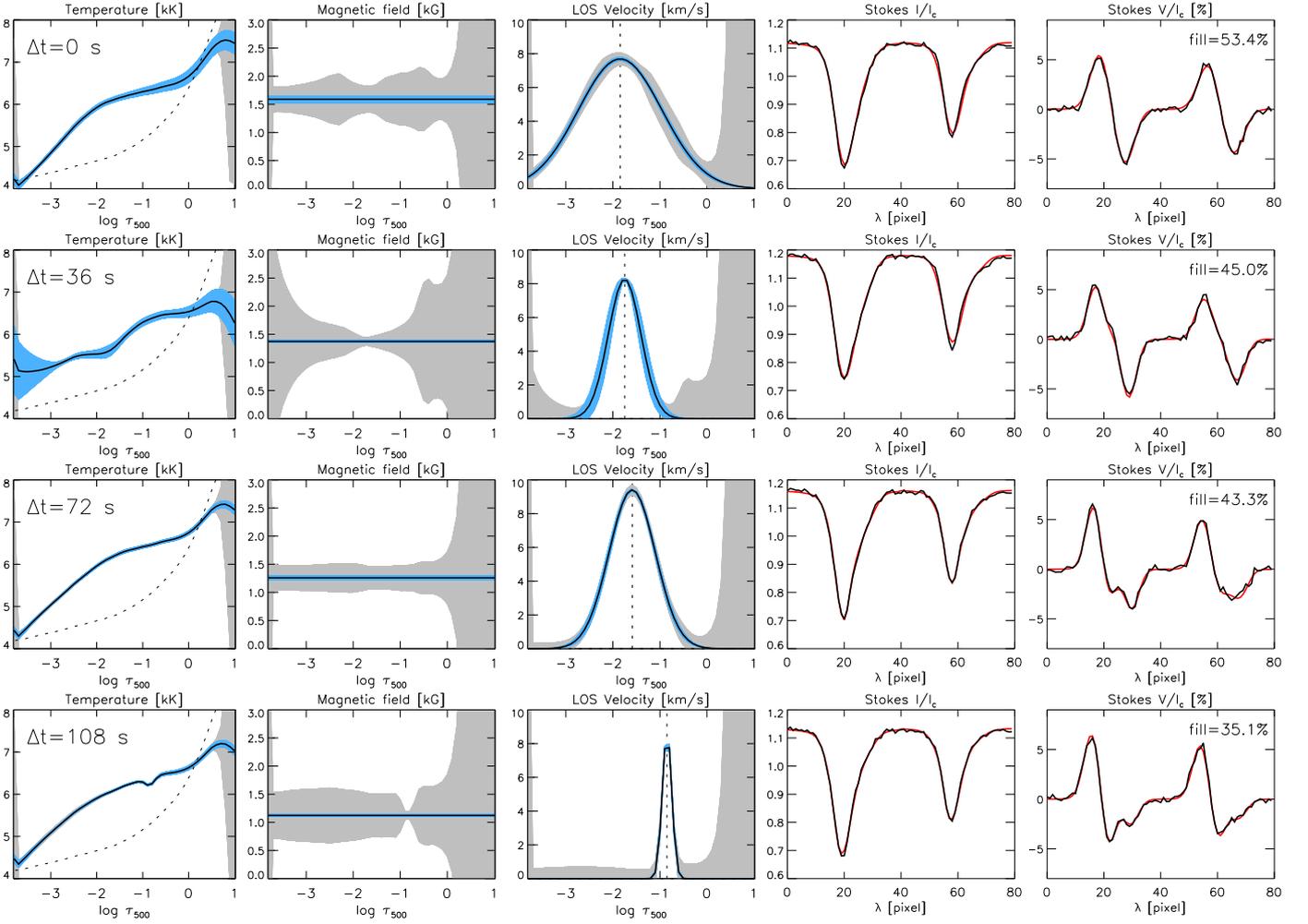}
\caption{Model atmosphere and Stokes profiles resulting from the SIRGAUSS inversion from one of the detected events. Each row corresponds to a different time running from top to bottom. The Stokes profiles are the same as presented by colored lines in Figure \ref{figper1}. The first column shows the temperature stratification. Second and third columns are the magnetic field and the LOS velocity stratifications.  The black line designates the mean atmosphere obtained from the Monte Carlo simulation, blue is the deviation from this mean solution and grey represents the non-sensitive regions (large grey regions indicate that the response of the Stokes profiles to perturbations in the model are very small). The last two columns show the observed Stokes $I$ and $V$ profiles in black and the fitted ones in red. The dashed line in the first column designates the HSRA reference atmosphere temperature and the vertical line in the third column marks the center of the Gaussian perturbation.}
\label{figsirgauss}
\end{figure*}

\subsection{Monte Carlo analysis}\label{MCsec}

The complexity of the atmosphere implies that the convergence to the most reliable solution is not granted. For this reason we made a statistical study to find the most probable solution that minimizes the $\chi^{2}$ parameter.  We performed 100 inversions with the same initial atmosphere. In each inversion we added a random noise to each Stokes parameter. The noise value is of the order of the real noise, i.e.\ $10^{-3} I_c$, and is different for each Stokes parameter. From the results of the 100 inversions we obtained solutions that correspond to a different $\chi^2$ minimum. We selected the most common solution for the location of the center of the Gaussian perturbation calculating the mode of the different results. Then we chose all the solutions that presented a Gaussian center position closer than $\pm$0.1 (in log $\tau$ units) to the most common perturbation center. The final solution of the inversion will be the mean value of all of these selected inversion results. The atmosphere stratifications examined in the following sections were always obtained using this Monte Carlo simulation.

\subsection{Inversion results}\label{SIRGAUSSsec}

The results from the inversion of the Stokes profiles examined in Section \ref{general_properties} are shown in Figure \ref{figsirgauss}. Each row shows a different time for the same pixel, running from top to bottom, corresponding to the four different colors marked in Figure \ref{figper1}. The first three columns show the atmospheric parameters of temperature, magnetic field and LOS velocity. The last two columns show the observed and fitted Stokes $I$ and $V$ profiles.

Black lines, in the three first columns, indicate the mean result from the Monte Carlo study. The standard deviations from these mean solutions are indicated in blue. Finally, grey depicts the area departure of each parameter that would produce a deviation equal to the noise in the Stokes profiles. This departure was obtained from the response functions of the atmospheric model to changes in the stratification of the atmospheric parameters. The last two columns show in black the real profile and in red the profile that comes from the mean atmosphere obtained from the Monte-Carlo study.

The temperature shows a similar behavior for the four snapshots, being much hotter than the HSRA reference temperature (dashed line) during the process. The magnetic field intensity is large, reaching values in the range 1--1.5 kG. It decreases slightly during the evolution of the bump through the center to the red wing of the line. The LOS velocity reproduces the same behavior we showed in Fig.~\ref{figsint}, a Gaussian perturbation that moves down into the photosphere from the top to the bottom layers.  Note that the center position of the Gaussian is also a free parameter in the SIRGAUSS code. The amplitude of the Gaussian perturbation reveals velocity values larger than 6 km/s, indicating that the plasma velocities involved in the process reach values over the estimated sound speed in the photosphere. We can also observe that the Gaussian perturbation goes down to deeper layers, as the bump is moving through the profile until it reaches the red wing. The last two columns show that the fitted profiles obtained from the inversion (red lines) perfectly match the observed profiles (black lines). Finally, the value of the filling factor decreases with time in this particular pixel although, as we will show later, this is not the typical behavior.

\section{Physical quantities outside hydrostatic equilibrium}\label{Outsideeq}

As a first approach, we performed the inversions assuming hydrostatic equilibrium (hereafter, HE). However, the results obtained in the previous section necessitated considering dynamical terms in the equation of motion because the LOS velocity stratification reaches values that are above the sound speed in the photosphere. The non-equilibrium nature of the atmosphere causes the HE gas pressure, gas density and geometrical heights to be incorrect. However, the error that these physical quantities produce in the inverted Stokes profiles is not very important, allowing us to change these physical quantities by a few orders of magnitude with a consequent change in the Stokes profiles in the order of the noise level \citep{Puschmann2010}. Consequently, we do not need to repeat the inversion outside the HE conditions because we would obtain the same temperature, magnetic field and LOS velocity stratifications. The aim of this section is to obtain the gas pressure, density and geometrical height stratifications compatibles with the equation of motion taking into account these high velocity gradients. In addition, this equation of motion also uses the optical depth-dependent temperature and LOS velocity stratification deduced from the inversions.

\subsection{The equation of motion}

In order to obtain a more accurate geometrical height scale where the Gaussian perturbation is moving down with time we need to solve the equation of motion \citep[see for instance][]{Priest1984}:

\begin{equation}
\rho\frac{D\vec{v}}{Dt}=-\vec{\nabla} P+\vec{j}\times \vec{B}+\rho\vec{g} \ ,
\label{eq1}
\end{equation}
with \vec{j} and \vec{g} the electrical current density vector and gravity, respectively. The rest of parameters are the gas density $\rho$ and gas pressure P, the magnetic field \textbf{B} and the velocity \textbf{v} vectors. In addition, the Lagrangian derivative can be defined by

\begin{equation}
\frac{D}{Dt}=\frac{\partial}{\partial t}+\vec{v}\cdot \vec{\nabla} \ .
\label{eq2}
\end{equation}

We are going to consider only the variations along the line of sight and the \textit{z} axis aligned with the LOS direction. For this reason the variables are only height ($z$) and time ($t$) dependent. The $z$-component of the equation of motion can be written as:

\begin{equation}
\frac{dP}{dz}+\rho g=-\rho(\frac{\partial v_z}{\partial t}+\frac{\partial \frac{1}{2}v_{z}^{2}}{\partial z})\ ,
\label{eq4}
\end{equation}
with $g$=+274 ms$^{-2}$.

The term $\vec{j}\times \vec{B}$ is neglected because we have supposed that the magnetic field is strictly vertical in the region where the event is detected. We cannot be sure of the accuracy of this assumption because the Stokes $Q$ and $U$ signals are always below the noise. However, the magnetic field being strong enough, its inclination should be small because otherwise we would have Stokes $Q$ and $U$ signal above the noise level. To prove that, we have synthesized the Stokes profiles using the atmospheres obtained in section \ref{SIRGAUSSsec} and we have found an upper limit for the magnetic field inclination of 10-15 degrees. If the inclination values are above this limit, the Stokes $Q$ and $U$ signals would be clearly distinguishable from the noise signal. Consequently, owing to the high signals of the Stokes $V$ profiles and the small area covered by these features we believe that the assumption of a vertical field concentrated inside an intergranular lane is reliable.

\subsection{Solution of the equation of motion}

We rewrite the equation of motion as a function of continuum optical depth at 500 nm $\tau_{500}$ in order to find an expression that preserves the optical depth dependence of the temperature and LOS velocity stratifications assumed in the inversions. We write the ideal gas equation neglecting radiation as

\begin{equation}
\rho=aP \ ,
\label{eq7}
\end{equation}
where $a$ \small  $= \frac{\mu}{RT}$ \normalsize, being $\mu$ the mean molecular weight. From the definition of optical depth we can write

\begin{equation}
\tau_{500} ln(10)dlog_{10} \tau_{500}=-\kappa \rho dz \ ,
\end{equation}
where $\kappa$ is the continuum absorption coefficient at 500 nm per gram. We then define $x=log_{10}\tau_{500}$ and, using equation \ref{eq7}, we obtain

\begin{equation}
\tau_{500}  ln(10)dx=-\kappa aPdz \ .
\end{equation}

Let us define a new variable $h$ as:

\begin{equation}
h=\frac{Pdz}{dx}=-\frac{\tau_{500} ln(10)}{\kappa a} \ .
\label{eqh}
\end{equation}

We can suppose that both parameters $h$ and $a$ do not depend on the gas pressure, i.e.\ they are only functions of the temperature. This is because the continuum absorption coefficient $\kappa$ has a small dependence on the gas pressure given by the Saha equation. We have checked that the variation of the  continuum absorption coefficient with temperature (\small  $\frac{d\kappa}{dT}$\normalsize) is many orders of magnitude larger than the variation of the continuum absorption coefficient with the gas pressure (\small$\frac{d\kappa}{dP}$\normalsize). For this reason, we can suppose that the continuum absorption coefficient is only optical depth dependent, i.e.\ $\kappa=\kappa(\tau_{500})$. On the other hand, we have calculated the compressibility coefficient $\alpha=\small\left( \frac{\partial \rm ln \rho}{\partial \rm ln P}\right) _{\tiny T}$\normalsize and we have found that $\alpha \simeq 1$ throughout the deep layers. As, from Eq. \ref{eq7}, $\alpha=1+ \left( \frac{\partial \rm ln \it a}{\partial \rm ln P}\right) _{\tiny T}$, we can assume that to first order $a=a(T)$. This result is produced by the low dependence of the mean molecular weight on gas pressure changes. In fact, a change of 1\% in the mean molecular stratification is obtained when we change the gas pressure by two orders of magnitude. Then, given that the temperature stratification obtained from the inversions does not depend on the gas pressure, i.e. $T=T(\tau_{500})$, $h$ and $a$ are only function of the optical depth.

The next step is to introduce into equation \ref{eq4} the definition of $dx$ from the first term of equation \ref{eqh} and the definition of $a$ using equation \ref{eq7}. After some algebra we obtain

\begin{equation}
\frac{dP}{dx}+A(x)-PB(x)=0 \ ,
\label{eq8}
\end{equation}
with

\begin{equation}
A(x)=ah(g+\frac{\partial v_z}{\partial t})
\end{equation}

\begin{equation}
B(x)=-a(\frac{d(\frac{1}{2}v_z^2)}{dx}) \ .
\end{equation}

The solution of the linear differential Equation \ref{eq8} is the following:

\begin{equation}
P(x)=e^{\int_{x_0}^{x}B(x)dx}\left[ P_0-\int_{x_0}^{x}A\left(x\right) e^{-\int_{x_0}^{x}B(x)dx}dx\right] \ .
\label{eq9}
\end{equation}

In order to solve equation \ref{eq9} we need to calculate the variation of the LOS velocity ($v_z$) with time (\small$\frac{\partial v_z}{\partial t}$\normalsize). We deduced this variation from the four different LOS velocity stratifications presented in Figure \ref{figsirgauss}. We analyzed the temporal variation of the entire LOS velocity stratification at every instant, obtaining the derivative of $v_z$ for each optical depth point of the vector. This term has a high uncertainty because we can only use four points to calculate it; the bump is only seen in four different instants in the Stokes $V$ profile owing to the limited cadence of the observation.

\begin{figure}

\centering

\hspace*{-0.3cm}\includegraphics[width=9.4cm]{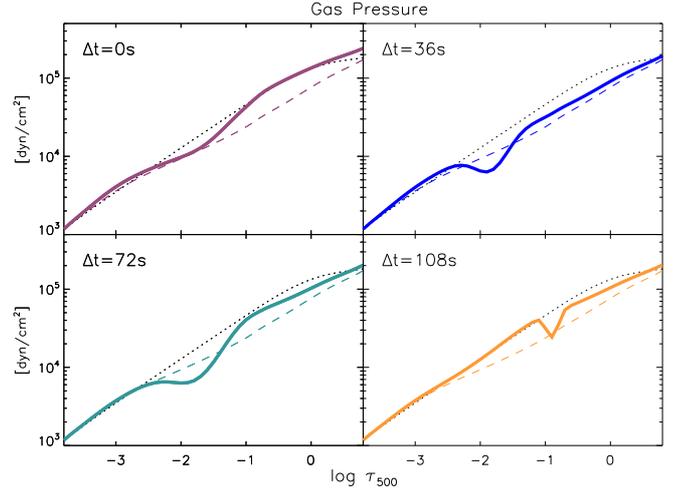}

\caption{Evolution of the gas pressure stratification outside HE assumption (solid). We maintain the color criteria used in Figure \ref{figper1}, purple for $\Delta t=0$, blue for $\Delta t=36$, green for $\Delta t=72$ and orange for $\Delta t=108$. Dashed lines are the solutions under HE and the dotted line is the gas pressure for the HSRA reference model.}

\label{figpg}

\end{figure}

\begin{figure}
\centering
\hspace*{-0.3cm}\includegraphics[width=9.4cm]{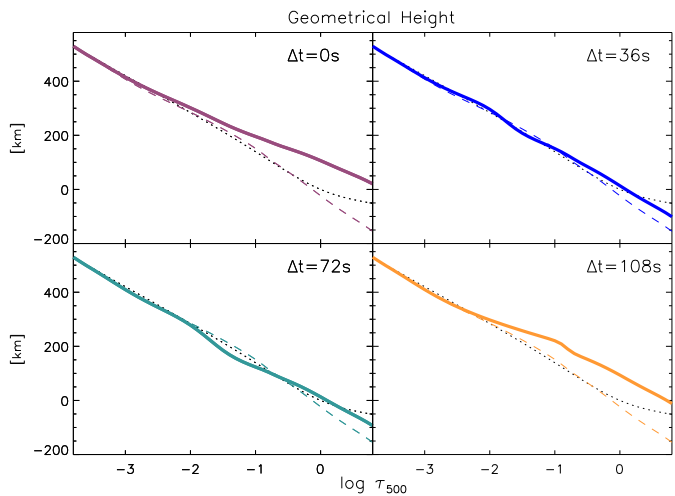}
\caption{Same as Figure \ref{figpg} for the relation between the optical depth and the geometrical height.}
\label{figzeta}
\end{figure}

\begin{figure*}
\centering
\hspace*{-0.6cm}\includegraphics[width=19.4cm]{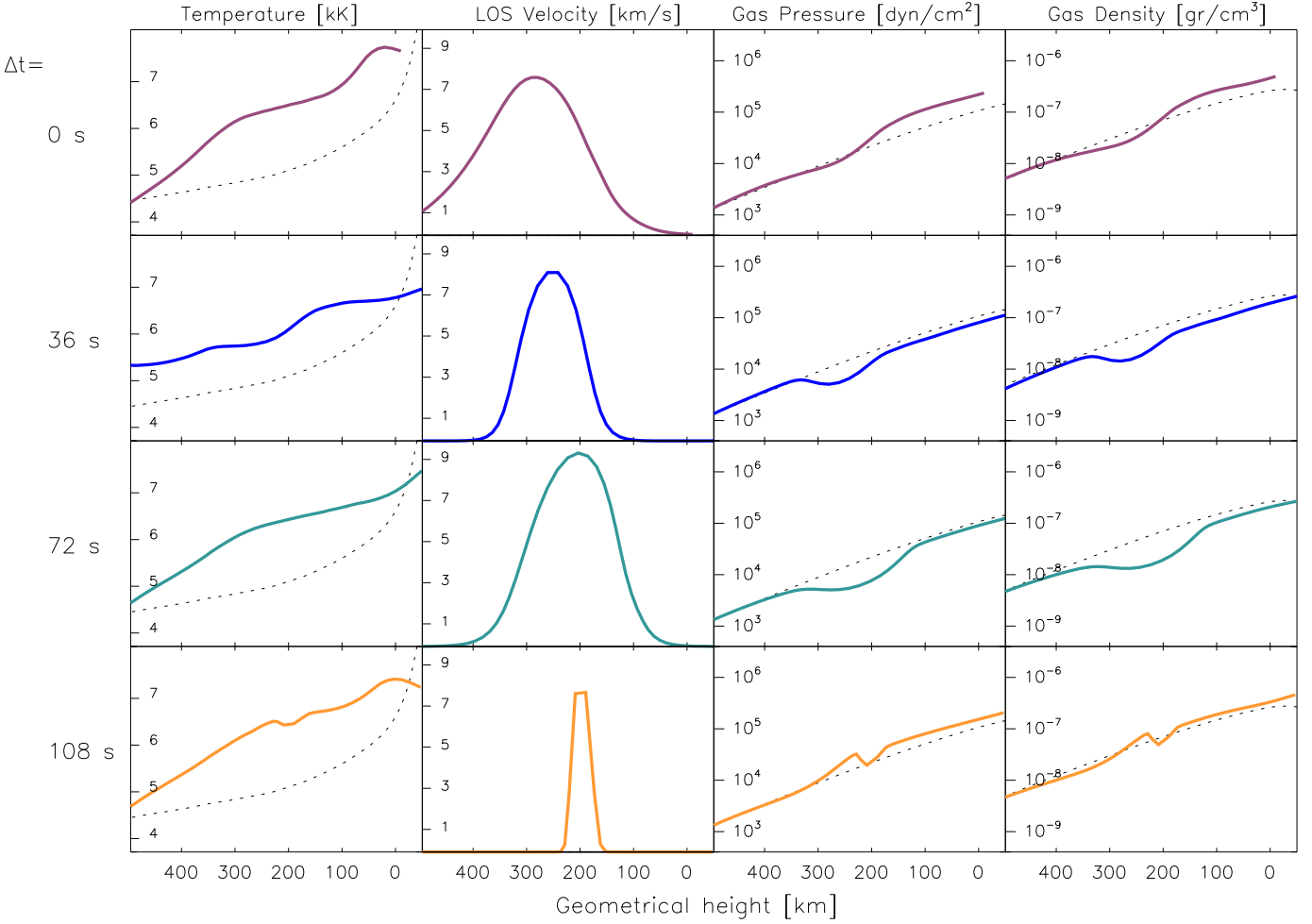}
\caption{Atmospheric parameters versus geometrical height. There are four columns with temperature, LOS velocity, gas pressure and density in solid lines. Dashed lines come from the HSRA reference model. The four rows correspond to the different time intervals defined in Figure \ref{figper1} with color lines. The geometrical heights from each row are different and come from those shown in Figure \ref{figzeta}.}
\label{fignoeq}
\end{figure*}

\begin{figure}
\centering
\hspace*{-0.3cm}\includegraphics[width=9.4cm]{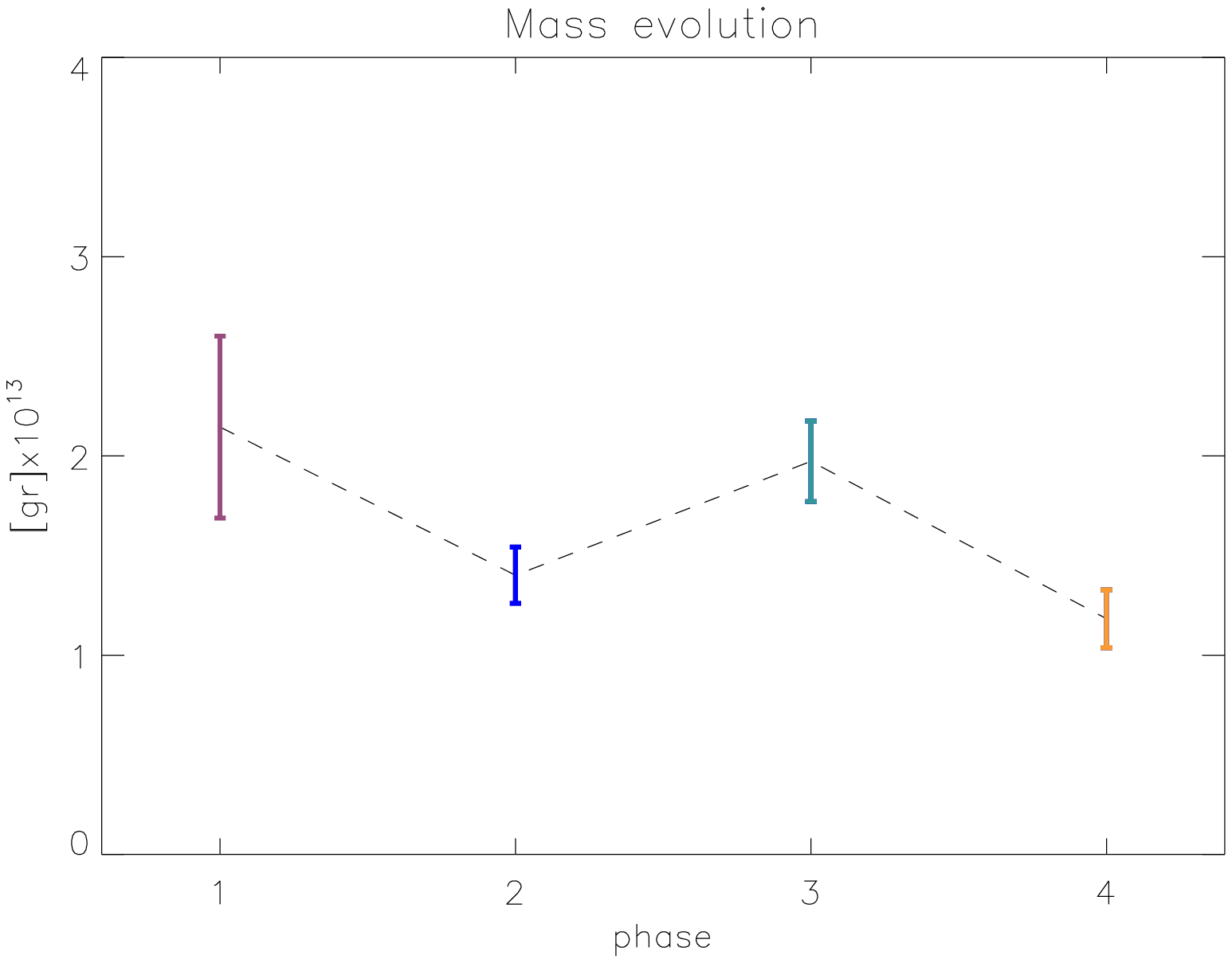}
\caption{Variation of the total mass during the event. There are four time steps that are the same as presented with colored lines in Figure \ref{figper1}.}
\label{figmass}
\end{figure}

\subsection{Gas pressure}

The results for the gas pressure stratification for the four different positions of the bump inside the Stokes $V$ profile, as marked by different colors in Figure \ref{figper1}, are shown in the panels of Figure \ref{figpg}. The dotted line is the gas pressure from the HSRA model, the dashed line represents the results from the HE assumption and solid lines correspond to the solution of the linear differential Equation \ref{eq9}. We have considered for the integration of this equation that $P_0$ (the gas pressure at $log\ \tau_{500}=-3.8$) has the same value as shown by the HSRA reference atmosphere model at the top boundary. This is why both lines, the reference atmosphere (dotted) and the gas pressure outside the equilibrium (solid), are joined together at the top of the atmosphere.

If we compare the results of the HE (dashed) and non-HE (solid) assumptions, we find that the main difference is the Gaussian perturbation that appears in the non-equilibrium gas pressure stratification. This perturbation is placed at the same location where the LOS velocity perturbation is; it also moves to deeper layers when the event evolves. Finally, the non-equilibrium gas pressure stratifications show almost the same behavior as the HSRA reference model at higher layers while they show slightly lower values than the HSRA reference atmosphere at the bottom of the photosphere.

\begin{figure}
\centering
\hspace*{-0.3cm}\includegraphics[width=9.4cm]{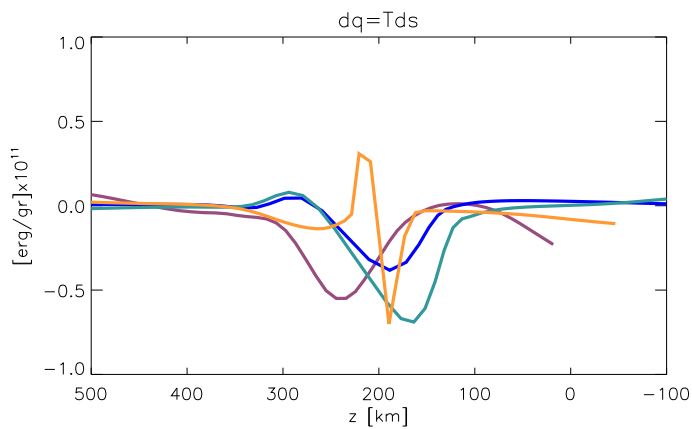}
\caption{Heat exchange that the perturbation undergoes during the process. We have used the same color code as in Figure \ref{figper1}.}
\label{figqm}
\end{figure}

\subsection{Geometrical heights}\label{GeometricalH}

Provided that in first order $h$ and $a$ are independent of the gas pressure, we can use Eq. \ref{eqh} to evaluate $h$ from the results of the inversions and to obtain the dependence of the geometrical height with optical depth using the gas pressure stratification. We do not have enough information to establish where the zero value of the geometrical heights is and we have chosen to put it at the top of the atmospheric stratification, at the same level as the HSRA reference model atmosphere. Figure \ref{figzeta} shows the geometrical height stratification outside hydrostatic equilibrium in solid lines while dashed lines are the result when we assume hydrostatic equilibrium. Again, the dotted line is the HSRA reference model. The geometrical height outside the HE assumption strongly diverges from the equilibrium results indicating that there is a large deviation in the relation between optical depth and geometrical heights due to the presence of highly speed magnetized plasma along the line of sight. This result implies that for the same optical depths we are sampling higher heights in the photosphere, and this is consistent with the fact that the perturbed models are less dense and more transparent than the unperturbed ones.

\subsection{Atmospheric parameters versus geometrical heights}

From the result of the previous section we can plot the atmospheric parameters inferred using the SIRGAUSS inversion versus geometrical height, obtained after taking into account the dynamical terms in the equation of motion. The results are shown in Figure \ref{fignoeq}. The first column shows the temperature stratification versus the geometrical height using the same color code as in Figure \ref{figper1}. The dashed line corresponds to the temperature stratification from the HSRA reference model. The second column shows the LOS velocity characterized by the Gaussian perturbation as shown in Figure \ref{figsirgauss}. The last two columns represent the gas pressure and density. Both magnitudes show a conspicuous negative Gaussian perturbation, located at the same height as the velocity perturbation, and a smooth behavior for the rest of the atmosphere stratification. Finally, the Gaussian perturbation shows a downward movement, as we found when we plotted the LOS velocity versus optical depth in Figure \ref{figsirgauss}. In addition, it also shows a width reduction during its evolution as if it were compressed when it moves towards denser layers. However, this behavior is clearly visible  in this example whilst it is not so evident for the rest of detected cases.

The total distance the center of the Gaussian has covered in 108 seconds is almost 200 km, indicating that the velocity of the center of the perturbation is nearly 2 km/s.

\subsection{Mass of the downward perturbation} \label{secmass}

The gas density shown in Figure \ref{fignoeq} reveals an inverse Gaussian perturbation moving downwards to deeper layers in the solar photosphere. We are going to analyze the mass value at the region where the inverse Gaussian perturbation is located to find out how it is changing during the process. In order to do this we are going to use the following equation:

\begin{equation}
\partial m=S\int\partial \rho dz \ ,
\label{eq10}
\end{equation}
where $\partial m$ is the mass inside the volume occupied by the Gaussian perturbation on the density stratification,  $dz$ the step in geometrical height and \small $S$ \normalsize the section occupied by the perturbation. This section \small $S$ \normalsize can be obtained as the pixel area multiplied by the filling factor of the magnetic component, \small $S=A f$\normalsize. The density $\partial \rho$ used to obtain the mass of the perturbation is the sum of all the density values from $-2\sigma$ to 2$\sigma$ with respect to the Gaussian center, $\sigma$ being the width of the Gaussian. We have plotted in Fig.~\ref{figmass} the results for the total mass for the same time evolution shown in colored lines in Figure \ref{figper1}. We can see the four mass values for the four evolution steps in the Stokes $V$ profiles. In order to estimate the uncertainty of these mass values we calculated the mass value for every different solution selected from the Monte Carlo analysis. The results presented in Fig.~\ref{figmass} indicate a decreasing in the mass as the perturbation moves to deeper layers. Although, if we take into account the large uncertainty of each point, the results could be compatible with a constant mass value during the whole process.

\subsection{Heat exchange}\label{heat}

The entropy variations of the perturbation could be obtained from changes in temperature and gas density as in \cite{Cox1968}:

\begin{equation}
dq=Tds=c_vdT-\frac{P\delta}{\rho^2\alpha}d\rho \ ,
\label{termo1}
\end{equation}
where $c_v$ is the specific heat at constant volume, $\alpha$ is the compressibility coefficient defined as $\alpha =\small \left(\frac{\partial ln \rho}{\partial ln P}\right)_T$\normalsize and $\delta$ is the thermal expansion coefficient given by $\delta =\small -\left(\frac{\partial ln\rho}{\partial lnT}\right)_T$\normalsize. The entropy and heat exchanges, $ds$ and $dq$, are defined per mass unit. In order to obtain the variation of the entropy of a mass element we need to introduce in Eq.\ \ref{termo1} the lagrangian variation of the gas density and temperature.

Figure \ref{figqm} shows the result of the heat exchange for the four different time steps with the same color code used in Figure \ref{figper1}. We can see that the four lines show the same behavior, presenting a smooth shape at the top and at the bottom of the atmosphere, negative values at the bottom part of the perturbation and small positive values at the top part of the perturbation. The explanation of this behavior could be that the first part of the perturbation is moving to deeper layers and is emitting radiation to the surrounding plasma that produces the detected negative heat exchange. At the same time, the last part of the perturbation is cooler than the plasma that it is leaving behind during its descent, so the heat exchange of the perturbation is positive because it is absorbing radiation from the external plasma. In all of these steps, the behavior is the same although the intensity of the exchange heat is increasing during the process. Finally, the consequence of the significant departure from adiabaticity (d\textit{s}=0) is that the radiative exchange of the perturbation with its surroundings has to be efficient and fast. In the next section we are going to obtain the radiative cooling time of the perturbation to determine whether this is feasible.

\subsection{Radiative cooling time}

To obtain an estimate of the time needed by a perturbation to exchange heat with its surroundings we followed the method used by \cite{Montesinos1993} for the analysis of syphon flows including radiative transfer. We assumed that the perturbation is under the optically thin limit. The radiative cooling time is then given by:

\begin{equation}
\tau_{rad}=\frac{c_p}{16\kappa \sigma T^3} \ ,
\end{equation}
where $c_p$ is the specific heat at constant pressure, $\sigma$ is the Stephan--Boltzmann constant and $\kappa$ the Rosseland opacity that could be approximated by:

\begin{equation}
\kappa=1.376\times10^{-23}p^{0.738}T^5 \ ,
\end{equation}
which units are in \small $cm^{2}/gr$\normalsize.

\begin{figure}
\centering
\hspace*{-0.3cm}\includegraphics[width=9.4cm]{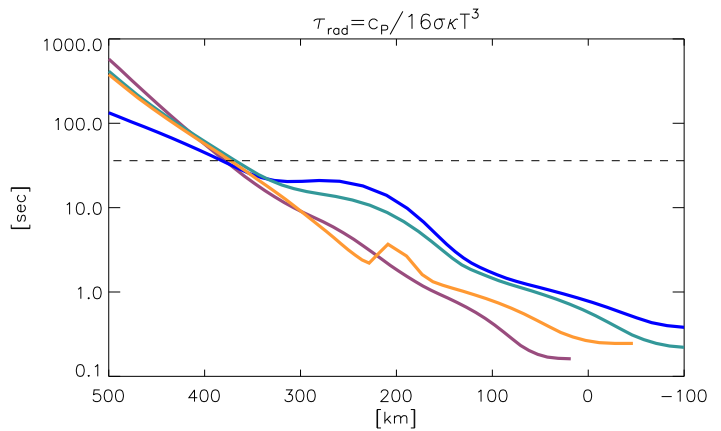}
\caption{Radiative cooling time for the Gaussian perturbation considering that it is optically thin. There are four colored lines with the same color code as used in Figure \ref{figper1}. The horizontal line represents the time cadence of the observation.}
\label{figtrad}
\end{figure}

\begin{figure}
\centering
\hspace*{-0.3cm}\includegraphics[width=9.4cm]{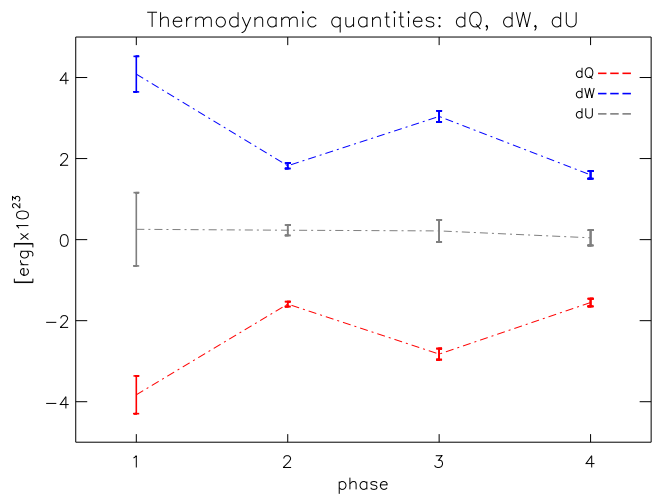}
\caption{Total heat, work and internal energy. The heat is represented in red, the work is shown in blue and the internal energy, as the sum of the two previous magnitudes, is represented in grey. There are four points that correspond to the different bump positions inside the Stokes $V$ profile as Figure \ref{figper1} shows.}
\label{figdiss}
\end{figure}

Figure \ref{figtrad} shows the radiative cooling time for the four different locations of the bump inside the Stokes $V$ profile. The gas pressure and temperature stratification used to obtain the cooling time are the same as shown in Figure \ref{fignoeq}. The horizontal line represents the time cadence of our observation, i.e.\ 36 seconds. The trace of the Gaussian perturbation is also present in the time stratification as a local time increase. All of the lines have a time value below the time cadence of the observation, where the Gaussian perturbation signal is located. Consequently, the time needed by the perturbation to reach equilibrium is much shorter than the time between observations, indicating that the perturbation is in thermal equilibrium with its surroundings.

\begin{figure*}

\centering

\includegraphics[width=18.5cm]{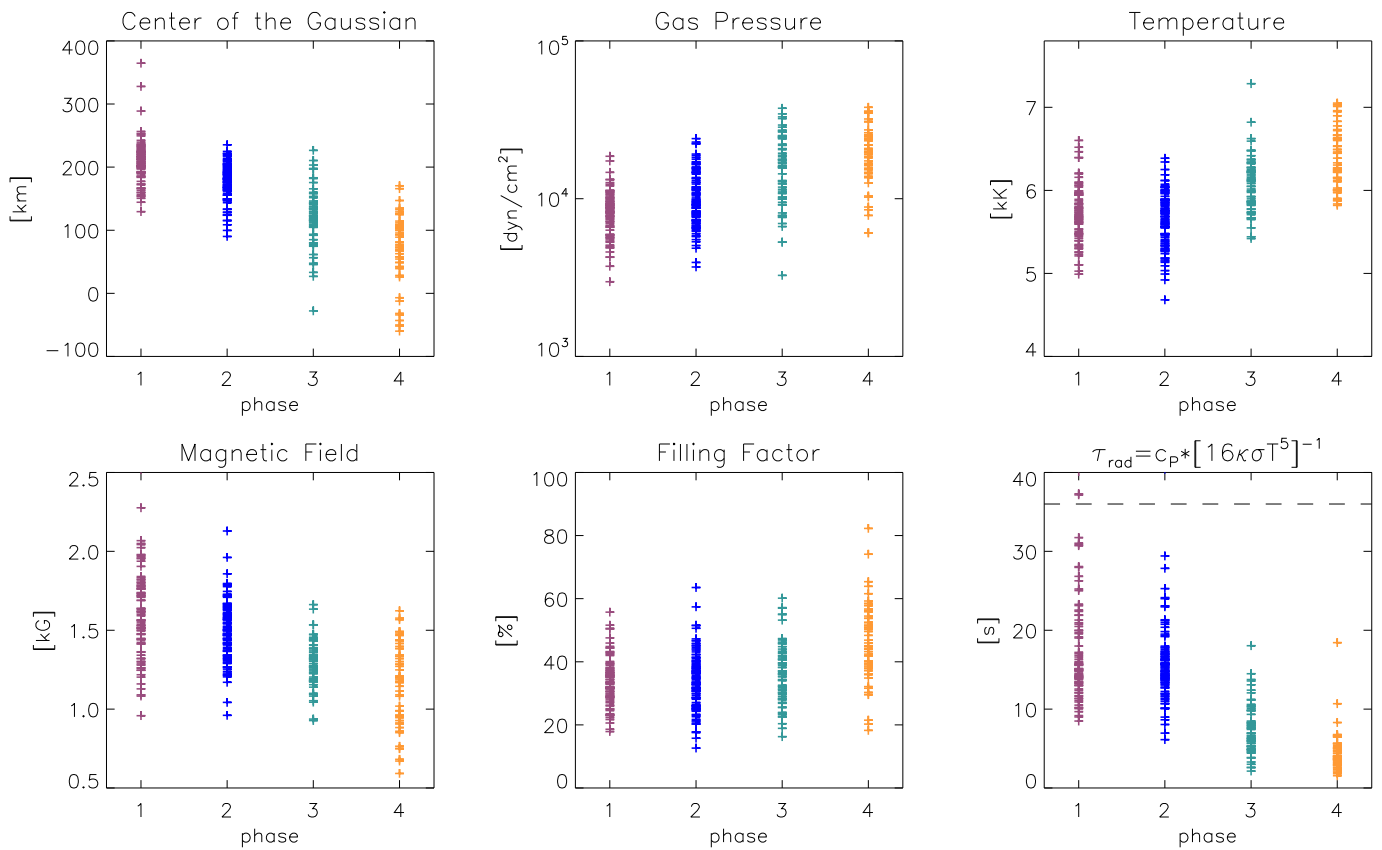}

\caption{Statistical results for the inversions of a large number of pixels selected from the maps presented in Figure \ref{figmagn}. We have used, in order, 92, 107, 69 and 57 pixels for each different phase presented with different colored lines in Figure \ref{figper1}. We have kept the same color code for each bump step. The first panel shows the location of the center of the Gaussian perturbation for each phase. The second top panel shows the gas pressure at the center of the Gaussian and the third top panel displays the temperature at the same location. The first panel of the second row presents the value of the magnetic intensity for each phase. The second bottom panel shows the filling factor of the magnetic component and the last panel displays the radiative cooling time for each phase.}

\label{figstatis}

\end{figure*}

\subsection{Total heat, work and internal energy}\label{TotalQWU}

The work done by external forces over the mass element can be defined as:

\begin{equation}
dw=\frac{P\delta}{\rho^2\alpha}d\rho \ ,
\end{equation}
and the total value of the work for every instant where the bump is visible inside the Stokes $V$ profile as:

\begin{equation}
dW=\int (dw) dm=\int (dw) S \partial \rho  dz \ ,
\end{equation}
where the section $S$, the increment in height $dz$ and the density perturbation $\partial\rho$ were described in Section \ref{secmass}.

To calculate the work we have repeated the same steps as in Section \ref{heat} to obtain the heat. Then we obtained the total heat $dQ=\small\int (dq) dm$\normalsize \ and the total work $dW$ integrating these quantities inside the volume occupied by the mass element. Finally, we calculated an uncertainty value for the total heat and total work in each time step using the different Monte Carlo solutions, see Sections \ref{MCsec} and \ref{secmass}.

The sum of both elements will give us the total internal energy of the perturbation. We show the evolution of these three thermodynamic parameters in Figure \ref{figdiss}. There are three colored lines with four points. The red line shows the behavior of the heat exchange by the perturbation with its surroundings. At the beginning, the mass element is dissipating heat at a high rate and when it starts to go down in the photosphere the heat exchange with the external medium is reduced. The blue line represents the work, which is positive at the beginning and is decreasing through the process, opposite to the heat exchange. The positive sign of the work means that the perturbation is suffering a compression because of the encounter with denser layers while it is moving downwards. Finally, the internal energy, as a sum of heat and work, is represented in grey. It is constant, inside the error bars, and close to zero because the pressure increment due to the encounter with a denser medium is nearly compensated with the heat radiated by the perturbation. This behavior indicates that the evolution of the mass element is under non-adiabatic conditions; all the energy gained by the compression of the downward mass motion is effectively radiated.

\section{Analysis of large number of events}

We have analyzed in detail a single event during the previous sections. In the present one, we are going to make a study of large number of events to check if the properties found and described before are the same for the remaining cases. In order to carry out this study we have selected pixels from the two \textit{normal} \textit{Hinode}/SP scan maps presented in Figure \ref{figmagn}. We chose the profiles inside the different redshifted patches placed over the whole map that shows the same spectral properties and time steps displayed with colored lines in Figure \ref{figper1}. After that, we inverted them using the same procedure followed in Sections \ref{Sirprocedure} and \ref{Outsideeq}. However, this time we performed the Monte Carlo analysis with only 20 inversions per profile. We also changed the method used to obtain the physical quantities outside the hydrostatic equilibrium hypothesis because we were examining the \textit{Hinode}/SP \textit{normal} maps. We could only use the variation of the atmospheric parameters with optical depth because we had no  information on the temporal evolution. We rewrote Eq. \ref{eq4} for this scenario as:

\begin{equation}
\frac{dP}{dz}+\rho g=-\rho\frac{\partial \frac{1}{2}v_{z}^{2}}{\partial z} \ .
\label{eq4mod}
\end{equation}

If we assume that the magnitude of the redshift of the bump in the Stokes $V$ profile is directly related to the temporal evolution, we can look for pixels that correspond to the different phases represented in Figure \ref{figper1}. We performed the classification by visual inspection and then we separated the results from the analysis of each phase to facilitate the visualization. We have found 92 pixels for the first bump step ($\Delta t=0$ in Fig.~\ref{figper1}), 107 pixels for the second bump step ($\Delta t=36$), 69 profiles for the third bump step ($\Delta t=72$) and 57 pixels for the last bump step ($\Delta t=108$). Figure \ref{figstatis} shows the physical parameters obtained from this amount of samples through the inversion of the Stokes profiles following the same method presented in Section \ref{Sirprocedure}. We show the results for the different atmospheric parameters in the region where the center of the Gaussian perturbation is located. The first top  panel shows the location of the center itself versus the geometrical height taking into account dynamical terms in the equation of motion. The cloud of points reveals that the center of the Gaussian is descending to deeper layers during its evolution. The top second panel shows the gas pressure at the same height where the center of the Gaussian is located. The pressure increases while the perturbation moves to denser layers. The top third panel corresponds to the temperature at the location of the Gaussian center. The temperature is increasing during its evolution while the perturbation is moving downwards and gets inside hotter plasma. This characteristic is in agreement with the total heat exchange presented in Fig.~\ref{figdiss}, where the heat exchange with the external medium decreases in the last steps because it is in equilibrium with the surrounding material. The first panel of the second row shows how the magnetic field intensity is slowly decreasing during the evolution of the perturbation. The following panel shows how the magnetic filling factor is slightly increasing with the evolution of the process. Because the magnetic field is diminishing while the filling factor is showing the opposite behavior, the resulting magnetic flux will be almost constant during the process. Finally, the last panel exhibits the radiative cooling time, at the center of the perturbation. Its value is small compared with the horizontal line that represents the cadence of the raster scan mode (36 seconds). This last property is in agreement with the results of the previous section that points to a non-adiabatic process.

In summary, the Gaussian perturbation is descending to deeper layers while it is emitting heat to the surroundings until it reaches an equilibrium state (see Figure \ref{figdiss}).

\section{Discussion}

The analysis of the distorted profiles detected by the \textit{Hinode}/SP instrument points to a possible mechanism that would originate in higher layers, probably at the bottom of the chromosphere, because we have found correlation with Ca~{\sc ii h} bright points.  The process then evolves and descends, leaving a detectable signal in Mg~{\sc i} b$_2$ Dopplergrams and, at the end, it reaches the low photosphere producing the strong polarimetric signal we have analyzed in the Fe~{\sc i} 6301.5 and 6302.5~\AA \ lines. However, after the careful and detailed analysis we have done, there is the important question to answer of which physical process produces these strong flows.

A possible mechanism that can produce strong downflows in the quiet-Sun photosphere is a syphon flow process, as studied by \cite{Rueedi1992} and \cite{Montesinos1993}. However, that mechanism seems inappropriate for explaining this process because we do not detect the other associated footpoint showing an upward velocity or the linear polarization signal that this loop configuration would produce in its top part. In fact, we only have observed a single polarity region with strong fields, before the appearance of the redshifted signal, during the process and after the disappearance of the peculiar Stokes $V$ profiles.

Another mechanism that could produce these features is a convective collapse process \citep{Parker1978,Webb1978,Spruit1979}. Photospheric layers are places prone to harbour a convective collapse event because that is where the horizontal granular flows lead to the concentration of enough magnetic flux in the intergranular lanes \cite[and references in]{Schussler1990} that would produce the convective instability. \cite{Shimizu2008} and \cite{Fischer2009} have suggested that the mechanism behind the downflow events they examined is convective collapse. This mechanism implies an intensification of the magnetic field after the strong downflows, but we have found the opposite behavior: the magnetic field decreases during the process. Furthermore, the convective collapse process usually ends with the formation of a stable magnetic flux tube \citep{Nagata2008}, but our results do not seem compatible with this scenario, at least in the iron line formation region, because almost all the detected events vanish after the disappearance of the strong redshifted signal. In addition, the destruction of the magnetic flux tube is sometimes related to a rebound of material that appears when the downward magnetized plasma meets denser layers. This rebound would produce an upward stream \citep{Grossmann1998} and a strongly blueshifted lobe in the Stokes $V$ profiles \citep{BellotRubio2001,SocasNavarro2005}. We did not find any trace of such blueshifted signals when the events vanish.

Another possibility we have examined is a magnetic reconnection \citep{Parker1963} between field lines in the upper photosphere that could produce opposite jets. One of them, descending,  would correspond with the features presented in this work (as we detect in the \textit{Hinode}/SP data) and the other, ascending, would produce the bright point in the Ca~{\sc ii h} spectral band that we have found. This hypothesis leads us to the physical mechanism behind the type {\sc ii} spicules.  Realistic MHD simulations proposed by \cite{Sykora2011} show that the origin of spicules involves dissipation of magnetic energy in non-adiabatic conditions. The main properties of this process are a reduction of the magnetic field strength, a downward shock and a heating in the atmosphere.  We have detected traces of these properties through the results of the inversion code. In addition, \cite{mcintosh2009} have observed  type {\sc ii} spicules in single polarity regions, which implies that the old conception of mixed polarity regions in the photosphere to produced a magnetic field reconnection is not always necessary.

However, we have no information on the chromosphere, so we can not presume that we have detected the disk center photospheric trace of the formation of spicules. But what we can do is compare with another chromospheric process: rapid blueshifted excursions (RBEs), first announced by \cite{Langangen2008} as the disk counterpart of type {\sc ii} spicules. They found a strongly blueshifted signal in the chromosphere measuring the Ca {\sc ii} line with high velocities of 15--20 km/s and short lifetimes of about 45 s. The authors concluded that the origin of the RBEs has to be some kind of magnetic reconnection due to the high energy released during the event. If a magnetic reconnection is happening and produces a very vertical and narrow blueshifted jet of material that is detected in the chromospheric layers, it seems plausible that the jet counterpart, narrow downward material, could be identified as the strong downflows we have examined in this work. They are very vertical and narrow, with lifetimes of 108 seconds and photospheric velocities near 10 km/s. In fact, the RBEs are characterized to be detected in regions close to strong magnetic fields, but not on top of them, as we have found for the redshifted events, see Figure \ref{figmagn}. At this point, with all of these similarities between both process, we believe that these strong downflows are the photospheric counterpart of the chromospheric RBEs, i.e.\ of type {\sc ii} spicules.

\section{Conclusions}

We have carefully examined the quiet-Sun strong downflows detected with \textit{Hinode}/SP as a redshifted Stokes $V$ signal far from the zero crossing point. These events are located in intergranular lanes and appeared where the magnetic field is strong. There is no evidence of opposite polarity regions or linear polarization signal patches close to them. After examining the time series, we establish a lifetime of 360$\pm$74 seconds and a mean size of 4$\pm$1.4 pixels. The analysis of the \textit{normal} maps reveals a rate of occurrence of 6$\times10^{-3}$ cases per arcsec$^{2}$.

The Stokes profiles present a characteristic polarimetric signal: Stokes $I$ shows a high continuum signal and line core values in comparison with a quiet-Sun profile from an intergranular lane region. Stokes $V$ profiles show a bump that in the first moment is present in the upper part of the line at the zero crossing wavelength. This profile evolves, presenting the same bump at different wavelength positions inside the line; in the red lobe and then in the red wing. Attending to the positions of the bump, the process evolution could be classified in four different steps. The lifetime of each step is lower than the cadence of the time series, i.e.\  36 seconds, because we always found a different step inside the pixel in consecutive images. The evolution of the wavelength position is always in the same direction, from the zero crossing point of the line to the red wing. We did not find the presence of this bump in the blue lobe of the Stokes $V$ profiles.

The strong Doppler-shifted signal detected in the lower photosphere using \textit{Hinode}/SP data was examined in higher atmospheric layers with the upper photospheric Mg~{\sc i} b$_2$ line and the lower chromospheric Ca~{\sc ii h} line. The Mg~{\sc i} b$_2$ Dopplergrams presented a patch of strong redshifted signal in the same place where the event is detected with the iron lines. At the same time, the Ca~{\sc ii h} spectral band images displayed a bright point at the same region. Both, Mg~{\sc i} b$_2$ and Ca~{\sc ii h} images, show the associated pattern before, or at the same time as, the detection of the strong downflows in the \textit{Hinode}/SP observation data but never after the detection of the process by the photospheric iron lines.

The qualitative behavior of Stokes $V$ profiles can be reproduced with models with a step in the LOS velocity stratification. In order to reproduce the Stokes $V$ amplitudes, with the Fe~{\sc i} 6301.5~\AA \ line intenser than the Fe~{\sc i} 6302.5~\AA \ line, we have chosen a hotter atmospheric model compared with the HSRA reference atmosphere. The bump evolution on the Stokes $V$ profiles can be obtained using a Gaussian perturbation in the LOS velocity that moves from the top to the bottom of the photosphere.

The atmospheric model compatible with the observed Stokes profiles has been inferred using the SIRGAUSS inversion code. The model sequence reveals a Gaussian perturbation in the LOS velocity moving downwards in the photosphere, embedded in a hot atmosphere. The magnetic field intensity is over kG values during the whole evolution and decreases slightly during the different phases.

The departure from hydrostatic equilibrium in the models is great because of the high velocities found, around 10 km/s. We obtained the gas pressure stratification taking into account the dynamical terms, $\frac{Dv}{Dt}=\frac{\partial v}{\partial t}+\tiny \vec{v}\cdot \vec{\nabla v}$\normalsize, in the equation of motion. The perturbation in the LOS velocity stratification produces a negative Gaussian perturbation in the gas pressure and density at the same region where the LOS velocity shows it.

The analysis of the perturbation heat exchange with its surroundings shows that the front part of the event is emitting heat to the outside medium, cooler than the perturbation, while the rear part is absorbing heat from the external media, hotter than the perturbation. This leads to a negative heat exchange in the front part of the perturbation and a positive heat exchange in the rear part. Assuming that the perturbation is optically thin, we employed the radiative cooling time definition used in the work of \cite{Montesinos1993} to ascertain how long the perturbation needs to reach an equilibrium state with its surroundings. We have obtained a time value lower than the time cadence of the observations, pointing to a strongly non-adiabatic process. We also study the evolution of the mass at the location of perturbation and we have found that it is nearly constant, within the error bar uncertainty, during the process.

If we assume that the magnitude of the redshift of the bump in the Stokes $V$ profile is directly related to the temporal evolution,  we can assume that the results of the study of the large number of cases found in the Hinode/SP normal map are compatible with the conclusions inferred from the time series observation. We have found that the center of the Gaussian perturbation is always moving downwards into the photosphere with time, while gas pressure and temperature at the center of the Gaussian perturbation increase with the evolution of the process. The magnetic field, chosen as constant with height, decreases with the evolution of the perturbation while the magnetic filling factor slightly increases, keeping the magnetic flux value almost constant. Finally, the radiative cooling time is always much lower than the lifetime of the event, confirming that it is a strongly non-adiabatic process.

The temporal variation of the perturbation displayed by the bump in the Stokes $V$ profiles, found in a large number of cases, is compatible with a \textit{plasmoid} of hot plasma that moves downward with high velocity. The different observed phases or steps correspond to the different optical depth locations of this \textit{plasmoid} during its descent. This perturbation descends in non-adiabatic conditions, undergoing a compression  by the denser external medium and effectively radiating  the generated heat due to this compression. This behavior causes the perturbation to remain always in thermal equilibrium (nearly constant internal energy). The magnetic flux confined inside the perturbation also shows a constant value during the process because the decrease in the magnetic field intensity is compensated by the increase in the filling factor that could be understood as an increase in the horizontal size of the perturbation.

All these results support the idea of a new process that takes place at the bottom of the chromosphere and evolves as a downward motion until the mid photosphere. We believe that this is the first time that the evolution of this event, through a detailed analysis of the Stokes profiles, is described. The most probable mechanism is a magnetic reconnection that is taking place at the top of the photosphere or the bottom of the chromosphere. In this sense, this process could be related to the formation of type {\sc ii} spicules. So far, we have no polarimetric information on the chromosphere. For this reason, to reveal the nature of these highly dynamics events, we need to wait for a possible simultaneous polarimetric observing modes of the chromosphere and the photosphere that the future major projects as EST, ATST and the space mission Solar-C would provide. Also, it would be necessary to examine realistic MHD simulations covering the chromosphere and the photosphere in order to look for similar events and to investigate the mechanism that produces them.

￼\begin{acknowledgement}
This work has been funded by the Spanish MINECO through Projects No. AYA2009-14105-C06-03, AYA2011-29833-C06-03 and AYA2012-39636-C06-06. The data used here were acquired within the framework of Hinode Operation Plan 14 (Hinode-Canary Islands joint campaign). \textit{Hinode} is a Japanese mission developed and launched by ISAS/JAXA, with NAOJ as a domestic partner, and NASA and STFC (UK) as international partners. It is operated by these agencies in cooperation with ESA and NSC (Norway).
￼\end{acknowledgement}

\bibliographystyle{aa} % style aa.bst

\bibliography{cqnbib.bib} % your references Yourfile.bib

\appendix

\section{Fe~{\sc i} 5250.2 Stokes profiles}

\begin{figure*}

\centering

\includegraphics[width=18cm]{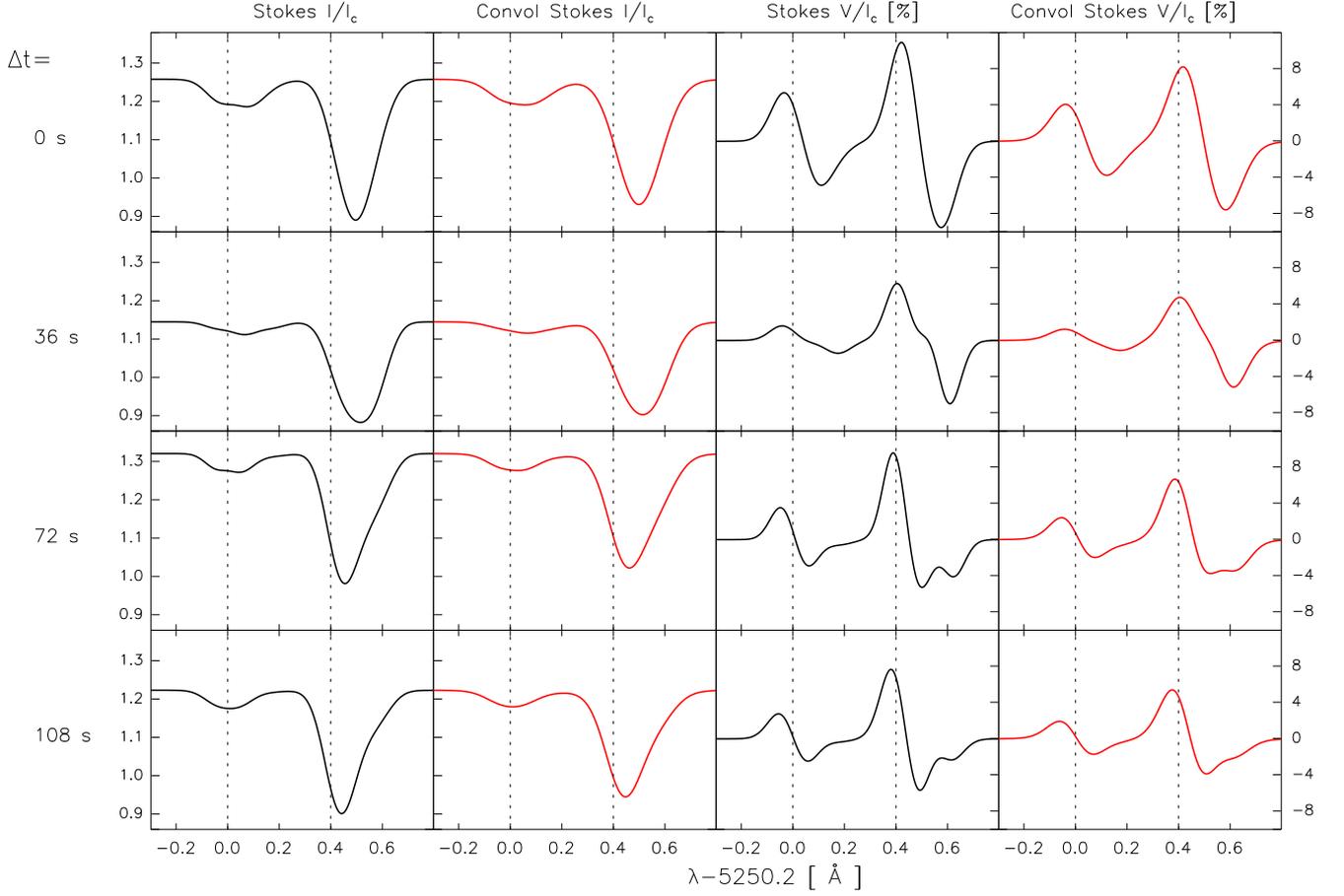}

\caption{Synthetic Stokes $I$ and $V$ profiles for the Fe~{\sc i} 5250.2 and 5250.6 \AA \ lines. We have used the atmospheres presented in Figure \ref{figsirgauss}. Black lines are the original synthetic profile while the red lines are the profiles after degrading them to the spectral resolution of IMaX. The dotted line designates the rest center of each line.}

\label{sintIMaX}

\end{figure*}

Inspired by the results of \cite{Borrero2010} who first analysed the quiet-Sun jets detected by Sunrise/IMaX and the possible detection of similar events by the later work of \citep{MartinezPillet2011a}, we synthesize the 5250 \AA \ spectral region (including the IMaX Fe {\sc i} line and the neighbouring Fe {\sc i} 5250.6 \AA \ line) using the atmosphere obtained in section \ref{Sirprocedure}.

Figure \ref{sintIMaX} shows the results of this synthesis. We only displayed Stokes $I$ and $V$ because the inclination angle was fixed to zero degrees during the inversion process. There are two different colors; black lines are for the synthetic profile while the red is that synthetic profile convolved with the Sunrise/IMaX spectral PSF.

Due to the high sensitivity of Fe~{\sc i} 5250.2 \AA \ line to the temperature, the Stokes $I$ parameter is very weak for these models. The second line, Fe~{\sc i} 5250.6 \AA, is less sensitive to the temperature and has a well defined shape with the red wing strongly bent due to the high velocity gradients inside the atmosphere.  The Fe~{\sc i} 5250.6 \AA \ line  has a higher line formation region and is less sensitive to the magnetic field. However, as happens with the Fe~{\sc i} 6301.5 and 6302.5 \AA \ lines measured by \textit{Hinode}/SP, the Stokes $V$ amplitude signal is higher and the bump is more prominent in the line that has a higher formation region. In fact, the bump has almost vanished in the line employed by the IMaX instrument, Fe 5250.2 \AA.

We can assume that these events, due to the strong temperature gradients and the large heights of formation in the photosphere, could not be detected by the line used by IMaX, even without the degradation effect produced by the spectral PSF of the instrument. We can see how the bump inside the profile of the second line has almost vanished, although in the last two steps it is still visible. We believe that the IMaX instrument could detect these events only if it is measuring the second line. For this reason we have to assume that the events analyzed by \cite{Borrero2010} and the studies that came later \citep{Borrero2012,Borrero2013,QuinteroNoda2013} examined a different physical process. In fact, they established that most of the cases were upflow events and were related with linear polarization patches.

\end{document}